\documentclass[aps,prx,twocolumn]{revtex4-2}

\pdfoutput=1

\usepackage{amsfonts,amsmath,amssymb}

\usepackage{graphicx}
\graphicspath{{graphics/}} 

\usepackage{xcolor}
\usepackage[colorlinks,citecolor=blue,linkcolor=blue,urlcolor=blue]{hyperref}
\usepackage{lipsum}  

\usepackage{lmodern} 
\usepackage[T1]{fontenc}
\usepackage[utf8]{inputenc} 
\usepackage{nicefrac}
\usepackage{textcomp}
\usepackage{braket}
\usepackage{slashed}
\usepackage{tensor}
\usepackage[subnum]{cases}
\usepackage{mathtools}
\usepackage{bm}
\usepackage{dsfont}
\usepackage{comment}

\newcommand{\abs}[1]{\left| #1 \right|}

\begin{document}

\title{Circuit QED theory of direct and dual Shapiro steps\\
with finite-size transmission line resonators}

\author{Federico Borletto}
\author{Luca Giacomelli}
\author{Cristiano Ciuti}
\affiliation{Université Paris Cité, CNRS, Matériaux et Phénomènes Quantiques, 75013 Paris, France}

\begin{abstract}
We investigate the occurrence of direct and dual Shapiro steps for a Josephson junction coupled to a finite-size transmission line resonator. We treat both problems through a circuit QED approach with a large, but finite number of photon modes. For the dual case, we do not assume the (approximate) charge-phase duality, but include the full multi-band dynamics for the Josephson junction. Mean-field equations within such Hamiltonian approach reproduce the result obtained through a dissipative classical equation when the number of transmission line modes is large enough. To account for quantum and thermal fluctuations, we go beyond the mean-field treatment within a truncated Wigner approach. The fluctuations are shown to modify both the direct and the dual steps.  We show how the dual steps are very sensitive to these fluctuations and identify the key physical parameters for the junction and the transmission line controlling their robustness, which is essential for applications to close the quantum metrological triangle. 
\end{abstract}

\maketitle

\section{Introduction}
The Quantum Metrological Triangle \cite{Likharev1985_th} provides a foundational framework for determining metrological standards through quantum mechanical phenomena. Central to this framework is the integer Quantum Hall Effect, which permits the accurate determination of the von Klitzing constant $R_\textnormal{K}=h/e^2 \simeq 25,812.807 \,\Omega$. This constant links the units of electrical potential (Volt) and electrical current (Ampère) through quantum mechanics.

Furthermore, the use of Josephson junctions \cite{Josephson1962} has allowed the observation of metrological features known as (direct) Shapiro steps \cite{Shapiro1963,Grimes1968}, obtained when a resistively shunted junction is biased with both a direct and an alternating current. These steps occur at integer multiples of the voltage $V_\textnormal{Q}=\Phi_0 \, \nu_\textnormal{AC}$, where $\Phi_0=h/(2e)$ is the flux quantum and $\nu_\textnormal{AC}$ represents the frequency of the applied Alternating Current (AC). Shapiro steps are the result of a synchronization phenomenon, typical of nonlinear systems. In particular, the nonlinear differential equations describing a Resistively and Capacitively Shunted Josephson Junction (RCSJ) in the overdamped limit can be mapped onto the circle map, a well-known model for mode locking \cite{Pikovsky2001}.
Direct Shapiro steps offer a quantum mechanical method for linking the Volt and the Hertz, and are at the basis of the Josephson voltage standard \cite{jeanneret2009application}.

Completing the metrological triangle by connecting the units of current and frequency remains a significant challenge. The main candidate phenomenon are dual Shapiro steps, that is steps at quantized values of current. These could be obtained by finding a device \textit{dual} to the Josephson junction, in which Cooper pair tunneling is replaced by phase slips, and is commonly referred to as \textit{quantum phase-slip junction}. Such a device can be obtained with superconducting nanowires \cite{Mooij2006}, or by using ultra-small Josephson junctions, that is junctions with a small capacitance, in a high-impedance environment \cite{averin1985bloch}.

This second possibility is actually an \textit{approximate} realization of a phase-slip element, in the sense that it does not exactly implement the constitutive relations  that are dual to the regular Josephson junction. Instead, the duality relies on a one-band projection of the Josephson junction spectrum, and on a sinusoidal approximation of such band \cite{Likharev1985_th, Averin1990, Guichard2010}.  Dual Shapiro steps in an ultra-small junction were observed in \cite{kuzmin1991observation}. This provided evidence for the existence of  the so-called Bloch oscillations in such junctions. However, the observation of clear steps in the current-voltage characteristic of the junction, and hence applications to metrology, has proven to be difficult \cite{kuzmin1994linewidth}.

Recent technological advancements have revived the interest in this problem, and important experimental steps have been taken in the last two years. Dual Shapiro steps in the IV characteristic were observed in \cite{Shaikhaidarov2022} by placing a nanowire in a highly inductive and resistive environment. A combination of large resistance and large (kinetic) inductance was also successful in observing the dual steps in small Josephson junctions \cite{kaap2024demonstration} (see also the related setup of \cite{Kaap2024}). A setup without a resistance was instead used in \cite{Crescini2023}, in which a small junction was inserted in a high impedance transmission line composed by an array of large junctions, leading to quantized current output at high voltage biases. While remarkable, all these experiments displayed plateaus that are less flat than those needed for metrological purposes, or modified by additional complex physics.

In this work, we focus on the class of physical systems used in \cite{Crescini2023}, that is one in which a Josephson junction is used and, instead of a resistive element, a transmission line resonator is used \cite{Weil2015, rastelli2018tunable, Lolli2015,PuertasMartinez2019Tunable,Manucharyan2019npj, Planat2019, FornDaz2016,  Kuzmin2021, Magazz2018, leger2019observation,Manucharyan2022, Manucharyan2023, Mehta2023, Kuzmin2023,leger2023revealing, Manucharyan2023arXiv, giacomelli2024emergent,Ashhab2024,Mukhopadhyay2023}
. Instead of taking the continuous limit of the Caldeira-Leggett model \cite{Caldeira1981,weiss2012quantum} of the environment, we approach it via a circuit QED treatment \cite{devoret1995quantum,girvin2014circuit,blais2021circuit} and consider a finite number of transmission line modes. With this, we aim at developing an exhaustive quantum description of direct Shapiro steps and their dual counterpart in these unconventional environments. This approach enables us to apply methodologies typical of many-body physics and quantum optics.

We first provide a description of direct Shapiro steps, for which we show that a mean-field approximation of the Heisenberg equations of motion for the junction and the environmental modes reproduces the result obtained with the classical RCSJ model in the limit of an infinite number of modes (vanishing mode frequency spacing). For what concerns the dual circuit, unlike conventional analyses, we do not simplify the model through a single-band approximation, and instead include the full quantum multiband dynamics together with the mean-field treatment. We show that the traditional conditions, which typically validate the use of the one-band approximation, may not be the optimal choice to observe dual steps, and that multi-band dynamics does not necessarily destroy the synchronization.

Going beyond these mean-field treatments, we include the effect of quantum and thermal fluctuations utilizing the truncated Wigner approximation (TWA) scheme \cite{Polkovnikov2010, Steel1998}. This approach extends our multi-band multi-mode mean-field theory, providing a framework that accounts for more complex and realistic scenarios. Crucially, the effect of the impedance of the environment is shown to be fundamental for the mean-field prediction to be robust against quantum fluctuations. Our work provides a novel approach to the problem of dual Shapiro steps, critically examining the approximations usually employed in theoretical works on the subject, with tools that are new to the field. Our results serve as a guide to experimental efforts towards the observation of metrologically useful dual Shapiro steps with Josephson junctions in transmission line environments.

The paper is organized as follows. In Section \ref{sec:shapiro} we develop a circuit QED description of Shapiro steps with a finite-size transmission line environment and recover the RCSJ result at the mean-field level when a large number of modes is considered. In Section \ref{sec:dual} we perform an analogous treatment of the dual circuit, with the added complexity of the multi-band dynamics. In Section \ref{sec:fluctuations} we develop a truncated Wigner description of both the direct and the dual cases to describe quantum and thermal fluctuations around the mean-field solution. We solve the problem using stochastic methods to assess the role of quantum dynamics on synchronization. In Section \ref{sec:conclusions} we draw our conclusions and pespectives.




\section{Theory of direct Shapiro steps}\label{sec:shapiro}

\subsection{Classical RCSJ model}
\begin{figure}[t!]
    \centering
    \includegraphics[width=1\columnwidth]{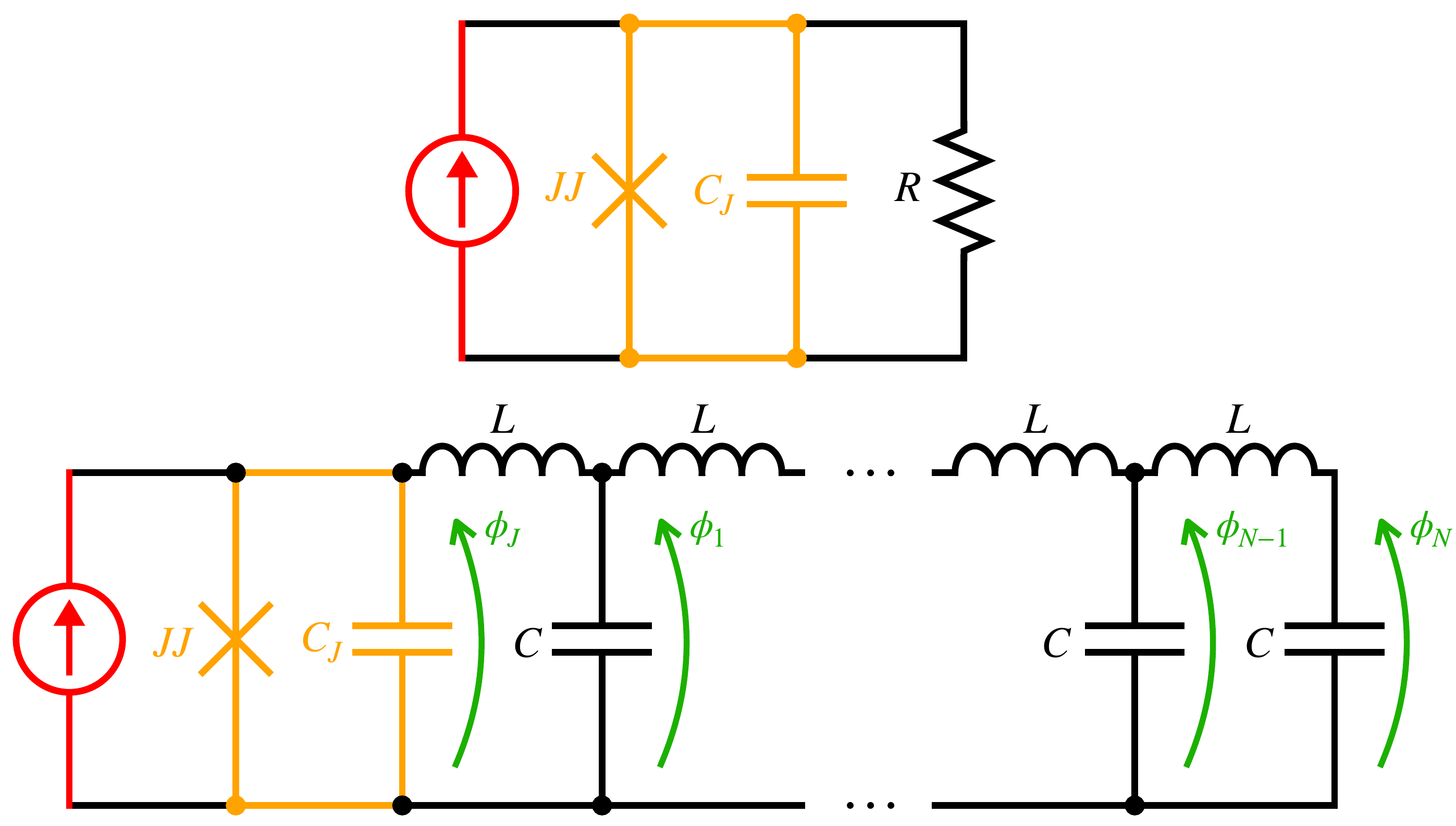}
    \caption{Upper panel: circuit diagram for the resistively and capacitively shunted junction (RCSJ) model. Lower panel: lumped-element representation of a finite-size transmission line terminated by an open circuit.}
    \label{fig:RCSJ}
\end{figure}

Direct Shapiro steps are the result of synchronization of the nonlinear dynamics of a Josephson junction with an external periodic driving. The simplest circuit giving rise to this phenomenon is the one depicted in the upper part of Fig.\ref{fig:RCSJ}, that is a parallel configuration with a tunneling element, a capacitor, and a resistor. This circuit is driven by a combined direct current (DC) and alternating current (AC) source.

The Shapiro steps represent the range of DC driving intensities over which synchronization occurs. This synchronization ensures that the DC component of the oscillating output voltage remains constant and is fixed to a multiple of the input frequency.

This is an effect that can be captured with a classical description of the Josephson phase by employing the resistively and capacitively shunted Josephson junction (RCSJ) model \cite{Martinis1990}, built on the established constitutive equations:
\begin{equation}
    \begin{cases}
        I_\textnormal{J}=I_\textnormal{C} \sin(\varphi_\textnormal{J}) \\
        V_\textnormal{J}=\frac{\hbar}{2e}\dot{\varphi}_\textnormal{J}  \, ,
    \end{cases}
\end{equation}
    where $I_\textnormal{C}$ represents the junction's critical current, and $\varphi_\textnormal{J}$ is the dimensionless flux (superconducting phase difference) across the junction. By applying Kirchhoff's current law at the upper node and introducing dimensionless variables, we derive the following dimensionless second-order differential equation (we drop the subscript $\textnormal{J}$ for the phase):
\begin{equation}\label{shap_diff_eq}
    \beta\frac{d^2 \varphi}{d\tau^2}+\frac{d \varphi}{d\tau}+\sin(\varphi)=\frac{I_\textnormal{DC}}{I_\textnormal{C}}+\frac{I_\textnormal{DC}}{I_\textnormal{C}} \sin(\bar\omega_\textnormal{AC} \tau) \, ,
\end{equation}
where we introduced $\tau=t/t_\textnormal{C}$ and the dimensionless AC driving frequency $\bar\omega_\textnormal{AC}=\omega_\textnormal{AC} t_\textnormal{C}$, with $t_\textnormal{C}=\frac{e}{\pi I_\textnormal{C}}\frac{R_\textnormal{Q}}{R}$ and $R_\textnormal{Q}=\frac{h}{4e^2}$ the quantum of resistance (it differs from the von Klitzing constant due to the double charge of Cooper pairs). Moreover, $\beta=\frac{\pi^2}{2}\frac{E_\textnormal{J}}{E_\textnormal{C}}\left(\frac{R}{R_\textnormal{Q}}\right)^2$ is the Stewart-McCumber parameter that quantifies the system's damping, with $E_\textnormal{J}=\frac{\hbar I_\textnormal{C}}{2e}$ the Josephson energy, and $E_\textnormal{C}=\frac{e^2}{2C}$ the capacitive energy.

The nonlinear differential equation \eqref{shap_diff_eq} can have a very complex dynamics. However, this is not the case if one works in the \textit{overdamped} regime characterized by $\beta\ll 1$, that can be obtained for small resistances. The steady state of the dynamics in such a regime, in combination with strong driving $I_\textnormal{AC}\gg I_\textnormal{C}$, can be predicted with analytical approaches \cite{Pikovsky2001}.

\begin{figure}[t!]
    \centering
    \includegraphics[width=1. \columnwidth]{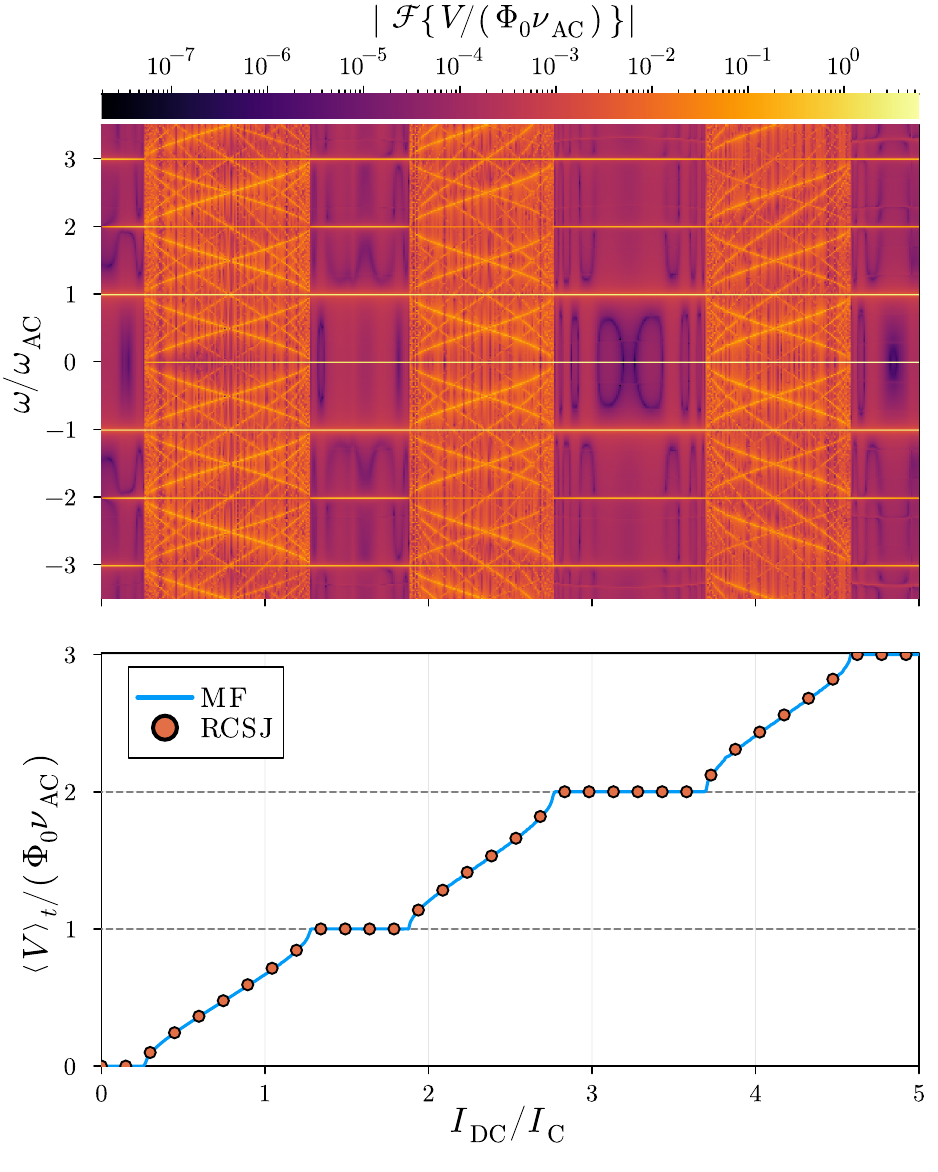}
    \caption{Bottom panel: Shapiro steps predicted by the mean-field dynamics for the junction coupled to a finite-size transmission line resonator, as described by the system of equations in (\ref{eom_shap_flux}).
    The circles depict the solution of the macroscopic RCSJ model with the same parameters for the Josephson junction, and with the resistance $R$ taken equal to the impedance $Z$ of the transmission line. Parameters: $E_\textnormal{J}/E_\textnormal{C}=1$, $Z/R_\textnormal{Q}=0.1$, $\hbar \omega_\textnormal{P}/E_\textnormal{C}=10$, $\hbar\Delta/E_\textnormal{C}=5 \times 10^{-3}$, $I_\textnormal{AC}/I_\textnormal{C}=5$, $\hbar\omega_\textnormal{AC}/E_\textnormal{C}=1$. The initial conditions for all variables of (\ref{eom_shap_flux}) are set to zero. The time averages are taken within this window over 50 periods of the AC source, excluding the first 25 to avoid transient effects.
    Top panel: logarithmic scale color plot of the absolute value of the Fourier transform of the output voltage versus current and frequency.}
    \label{fig:direct_shapiro_MF}
\end{figure}

More in general, one can numerically solve \eqref{shap_diff_eq} in the aforementioned regime to obtain the flux and then compute the time average of the output voltage in the steady state. As shown by the circle markers in the bottom panel of Fig.\ref{fig:direct_shapiro_MF}, neat plateaus occur for $\langle V \rangle_t=k\Phi_0 \nu_\textnormal{AC}$, where $k$ is an integer, $\Phi_0=\frac{h}{2e}$ the flux quantum, and $\nu_\textnormal{AC}$ the AC driving frequency.


\subsection{Circuit QED theory with a finite-size transmission line resonator}

To analyze this phenomenon at a quantum level, we adopt a parallel configuration similar to the classical RCSJ model, substituting the resistor with a lumped-element transmission line, as shown in the lower panel of Fig.\ref{fig:RCSJ}. This setup yields a Caldeira-Leggett model, which displays genuine dissipative behavior solely in the limit of infinite size. In contrast, the systems that we consider in our calculations are finite-size and conservative.
As detailed in Appendix \ref{appendixA}, we apply the circuit QED methodology \cite{devoret1995quantum} to derive the circuit Lagrangian in terms of the fluxes defined in Fig.\ref{fig:RCSJ}, diagonalize the environmental modes and perform a Legendre transform to obtain the corresponding Hamiltonian. 
The key parameters of the transmission line are the characteristic impedance $Z=\sqrt{L/C}$, the plasma frequency $\omega_\textnormal{P}=2/\sqrt{LC}$ and the free spectral range (i.e. the frequency mode spacing) $\Delta=\frac{\pi}{2}\frac{\omega_\textnormal{P}}{N}$, where $N$ is the number of LC oscillators. The modes operators and junction observables are then canonically quantized to obtain the quantum Hamiltonian in the \textit{flux gauge} (see Appendix \ref{appendixA}):
\begin{equation}
\begin{split}\label{ham_shap_flux}
\hat{H}_1 &= \sum_{i=1}^{N} \hbar \omega_i \hat{a}^\dagger_i \hat{a}_i+4E_\textnormal{C}\hat{n}^2_\textnormal{J} -E_\textnormal{J} \cos(\hat{\varphi}_\textnormal{J}) \\ 
    &+\left(\frac{\hbar}{2e} \right)^2\frac{\hat{\varphi}^2_\textnormal{J}}{2L} -\hat{\varphi}_\textnormal{J} \sum_{i=1}^{N} g_i (\hat{a}_i+\hat{a}^\dagger_i) -\frac{\hbar}{2e}I(t)\hat{\varphi}_\textnormal{J} \, .
\end{split}
\end{equation}
Here $\hat a_i$ are the bosonic mode operators, $\hat n_\textnormal{J}$ and $\hat \varphi_\textnormal{J}$ are the junction's charge and phase operators, and $I(t)$ is the current drive. Moreover, $\omega_i=\omega_\textnormal{P} \left| \sin(k_i/2)\right|$ is the frequency of the $i$-th mode with $k_i=\pi \frac{2i-1}{2N+1}$ and $i\in [1,\dots,N]$. The corresponding coupling to the junction is quantified by the flux-gauge coupling constant: 
\begin{equation}
g_i=\frac{1}{\sqrt{8\pi}}\left(\frac{R_\textnormal{Q}}{Z} \right)^{1/2}\frac{\hbar \omega_\textnormal{P}}{(2N+1)^{1/2}}\frac{\sin(k_i)}{\left|\sin(k_i/2) \right|^{1/2}} \, .
\end{equation}
Note that the term quadratic in $\hat \varphi_\textnormal{J}$ in Eq. \eqref{ham_shap_flux} is the gauge-dependent Caldeira-Leggett \textit{counterterm} \cite{weiss2012quantum}, whose prefactor obeys the \textit{sum rule} $\frac{E_\textnormal{L}}{2}\equiv\left( \frac{\hbar}{2e}\right)^2\frac{1}{2L}=\sum_{i=1}^{N} \frac{g_i^2}{\hbar\omega_i}$.

Given the Hamiltonian, we can derive the Heisenberg equations for the time-dependent operators:
\begin{equation}\label{eom_shap_flux}
    \begin{cases}
    \dot{\hat{a}}_i=-i\omega_i\hat{a}_i+\frac{i}{\hbar}g_i \hat{\varphi}_\textnormal{J}\\[10pt]
    \dot{\hat{\varphi}}_\textnormal{J}=\frac{8E_\textnormal{C}}{\hbar}\hat{n}_\textnormal{J} \\[10pt]
    \dot{\hat{n}}_\textnormal{J}=-\frac{E_\textnormal{L}}{\hbar}\hat{\varphi}_\textnormal{J}-\frac{E_\textnormal{J}}{\hbar}\sin(\hat{\varphi}_\textnormal{J})+\frac{1}{\hbar}\sum_i g_i(\hat{a}_i+\hat{a}^\dagger_i)+\frac{I(t)}{2e} \, .
    \end{cases}
\end{equation}

\subsection{Mean-field approximation and numerics}
In principle, the system \eqref{eom_shap_flux} describes the response to driving of the whole setup, i.e. junction plus transmission line. It is however impractical to solve them exactly, and this kind of Caldeira-Leggett representation of the environment is usually used as a starting point for analytical treatments in the continuum limit, or to derive effective master equations for a part of the global system. The system under study is however difficult to approach with standard Lindblad-like master equations because of the intrinsically ultrastrong multimode coupling of the junction with the environment, especially in the over-damped low-impedance regime we are interested in.

Therefore, we first tackle this dynamical problem at the mean-field level (we will go beyond this approximation in Section \ref{sec:fluctuations}), that is we solve the equations in which operators are replaced with complex numbers. This is equivalent to taking the average of the equations on an Ansatz state of the form $\ket{\psi}=\bigotimes_i \ket{\alpha_i}\otimes \ket{\varphi}$, with $\ket{\alpha_i}$ being a coherent state for mode $i$, and $\ket{\varphi}$ an eigenstate of the junction operator $\hat{\varphi}_\textnormal{J}$. With this Ansatz, we are left with a set of $N+2$ differential equations to solve. 

To obtain a point of the current-voltage characteristics, we numerically solve these differential equations with the combined DC and AC driving, taking zero initial conditions for all variables. Given the fact that we are treating a closed system, the evolution will not reach a steady state and the transmission line has a return time fixed by the free spectral range $t_\textnormal{R}=1/\Delta$. To perform the time average of the output voltage, we restrict to times smaller than $t_\textnormal{R}$, but large enough to avoid the initial transient dynamics. In practice, the transmission line should be long enough, so that $t_\textnormal{R}$ is larger than the time over which the output is measured \cite{Pekola2024}.

In the bottom panel of Fig.\ref{fig:direct_shapiro_MF} we show with a solid line the result of this computation, repeated for different amplitudes of the DC driving. The parameters are taken equal to the ones used for the RCSJ model; in particular, the characteristic impedance of the transmission line is set to the same value as the resistance in the classical model. One can see that the obtained IV curve has exactly the same shape as the previous one. Hence, a mean field approximation of the Heisenberg equations \eqref{eom_shap_flux} for the transmission line reproduces the RCSJ result. Notice that a large number of modes is needed to have a long enough integration time to obtain a clean IV curve.

It is instructive to consider the Fourier transform of the output voltage dynamics for different values of the DC driving amplitude. This quantity is reported in the upper panel of Fig.\ref{fig:direct_shapiro_MF}. This shows how Shapiro steps are rooted in a synchronization mechanism. In fact, while outside the plateaus there is a complex spectral structure, inside the steps the only relevant components of the Fourier transform are the driving frequency and its higher harmonics.


\begin{figure}[t!]
    \centering
     \includegraphics[width=\columnwidth]{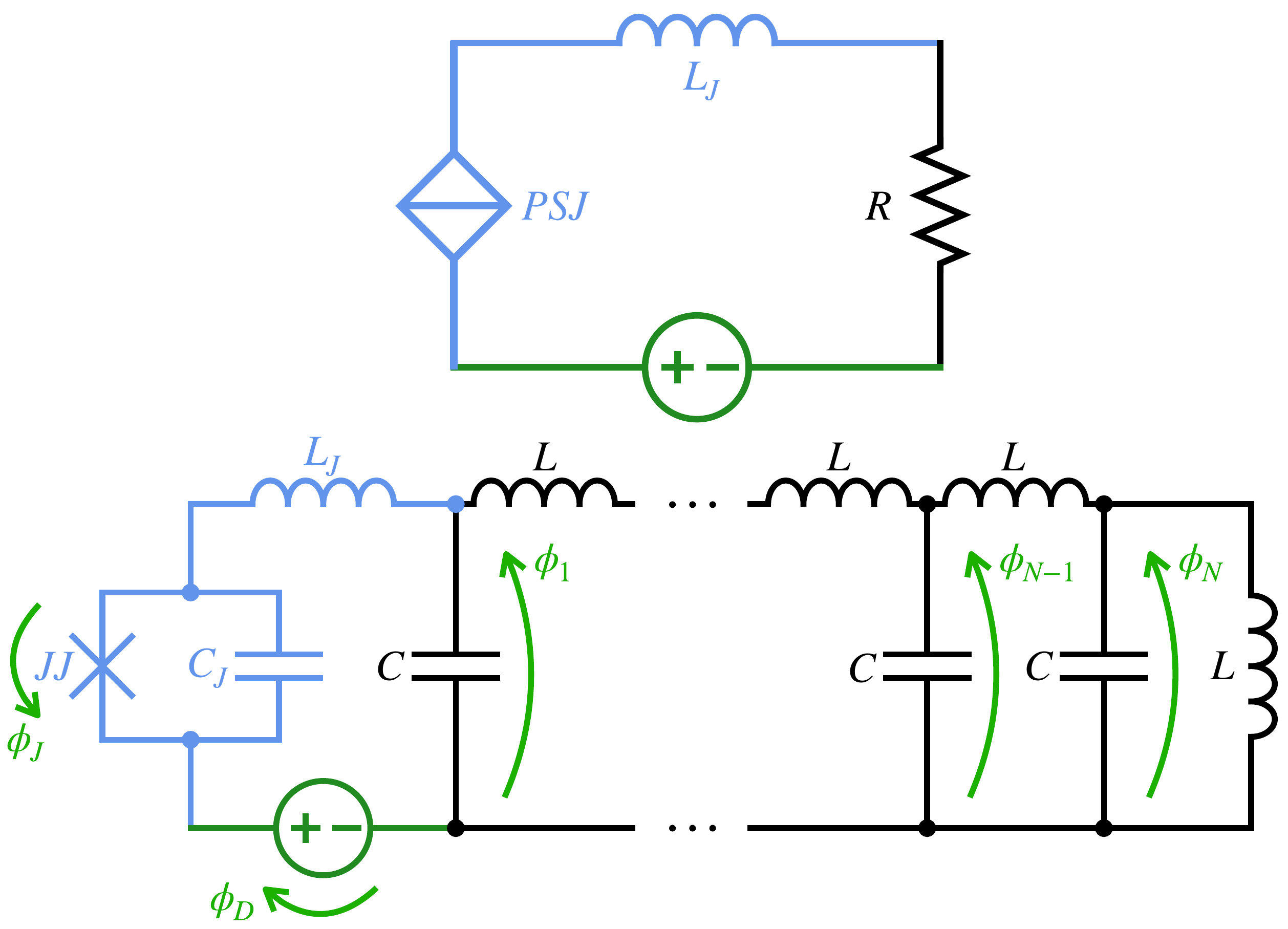}
    \caption{Top panel: circuit model with a phase-slip Junction (PSJ). Bottom panel: microscopic circuit, where the resistor has been replaced by a finite-size transmission line in the lumped-element representation. The ideal phase-slip component is replaced by a capacitively shunted Josephson junction terminated by a short circuit.}
    \label{fig:dual_circuit}
\end{figure}

\section{Theory of dual Shapiro steps}\label{sec:dual}

\subsection{Macroscopic model with non-linear capacitor}
We now examine the more complicated case of dual Shapiro steps. From the perspective of circuit theory, a circuit dual to the RCSJ model can be defined. The parallel configuration becomes a series configuration, the current source is replaced by a voltage source, and the capacitor is replaced by an inductor. As for the Josephson junction, a dual circuit component known as the \textit{phase-slip junction} (PSJ) is needed, and the constitutive relations it obeys are:
\begin{equation}\label{dual_const_rel}
    \begin{cases}
        I_\textnormal{PSJ}=\dot{Q}_\textnormal{PSJ} \\
        V_\textnormal{PSJ}=V_\textnormal{C} \sin(2\pi n_\textnormal{PSJ}) \, .
    \end{cases}
\end{equation}
In other words, if the constitutive relations of a Josephson junction can be seen as those of a nonlinear inductor, a phase-slip junction behaves as a nonlinear capacitor.

The macroscopic circuit for dual Shapiro steps is depicted in the upper part of Fig.\ref{fig:dual_circuit}. By applying Kirchhoff’s voltage law to the loop and adopting appropriate dimensionless variables, we derive an equation identical in form to (\ref{shap_diff_eq}) (we drop the subscript PSJ for better readability):
\begin{equation}\label{dual_diff_eq}
    \tilde{\beta}\frac{d^2 n}{d\tau^2}+\frac{d n}{d\tau}+\sin(2\pi n)=\frac{V_\textnormal{DC}}{V_\textnormal{C}}+\frac{V_\textnormal{AC}}{V_\textnormal{C}} \sin(\tilde\omega_\mathrm{AC} \tau) \, .
\end{equation}
Here we have $\tau=t/\tilde{t}_\textnormal{C}$ and the dimensionless AC driving frequency $\tilde\omega_\textnormal{AC}=\omega_\textnormal{AC} \tilde{t}_\textnormal{C}$, with $\tilde{t}_\textnormal{C}=\frac{\pi \hbar}{e V_\textnormal{C}}\frac{R}{R_\textnormal{Q}}$. Moreover, $\tilde{\beta}=\frac{1}{2\pi}\frac{U_0}{E_\textnormal{L}}\left(\frac{R_\textnormal{Q}}{R} \right)^2$ is the Stewart-McCumber parameter for the dual case, $E_\textnormal{L}=\left( \frac{\hbar}{2e}\right)^2\frac{1}{L_\textnormal{J}}$ the inductive energy, and $U_0=\frac{e V_\textnormal{C}}{\pi}$ the energy of the phase-slip junction. The phenomenology will be the same as the direct case, hence we expect to observe plateaus in the $\langle I\rangle_t$ versus $V_\textnormal{DC}$ characteristics whenever $\langle I\rangle_t=2ek \nu_\textnormal{AC}$, with $k \in \mathbb{Z}$. Notice how the overdamped regime for the charge dynamics $\tilde\beta\ll 1$ is obtained at high resistance $R\gg R_\textnormal{Q}$.

Quantum phase-slip junctions can be obtained with superconducting nanowires \cite{Mooij2006}, that directly exhibit the dual constitutive relations (\ref{dual_const_rel}). Here we focus instead on Josephson junctions once again, as they can \textit{approximately} behave as a quantum phase-slip junction. We discuss this approximate duality in detail after the derivation of the circuit QED Hamiltonian.
   

\subsection{Circuit QED theory of dual Shapiro steps}

Analogously to the direct case, we consider a circuit in which the resistance is replaced by a transmission line resonator, as depicted in the lower part of Fig.\ref{fig:dual_circuit}. Notice that this time the transmission line is terminated with a short circuit to maintain a loop configuration 
(and to obtain the dual of the original transmission line). As we already did for the previous case, we can write the Lagrangian of this circuit, and then quantize the corresponding Hamiltonian (details in Appendix \ref{appendixA}). This procedure yields the flux-gauge Hamiltonian:
\begin{equation}\label{ham_dual_flux}
\begin{split}
    &\hat{H}_2=\sum_{i=1}^N\hbar \tilde\omega_{i}\hat{a}^\dagger_i \hat{a}_i+
    4E_\textnormal{C} \hat{n}_\textnormal{J}^2 -E_\textnormal{J} \cos(\hat{\varphi}_\textnormal{J})+\frac{1}{2}E_\textnormal{L} \hat{\varphi}_\textnormal{J}^2+ \\ +&\hat{\varphi}_\textnormal{J} \left[\sum_{i=1}^N \tilde g_i (\hat{a}_i+\hat{a}^\dagger_i)-E_\textnormal{L}\varphi_\textnormal{D}(t)\right] -\varphi_\textnormal{D}(t)\sum_{i=1}^N \tilde g_i (\hat{a}_i+\hat{a}^\dagger_i) \, .
\end{split}
\end{equation}
Here $\varphi_\textnormal{D}$ is the time-dependent (classical) flux associated to the voltage drive (the time integral of this voltage). In this case the mode frequencies $\tilde\omega_i$ and the couplings $\tilde g_i$ are computed numerically, since there is no exact analytical expression for a finite-size transmission line due to the coupling inductance $L_\textnormal{J}$. 

The behavior of the Josephson junction as a phase-slip junction emerges once one considers the Josephson Hamiltonian in terms of Bloch bands, namely
\begin{equation}
\hat{H}_\textnormal{J}=4E_\textnormal{C} \hat{n}_\textnormal{J}^2-E_\textnormal{J}\cos(\hat{\varphi}_\textnormal{J})=\sum_m \int dq\, \varepsilon_m(q)\ket{m,q}\bra{m,q},
\end{equation}
where $\braket{\varphi | m,k}=e^{ik\varphi}u_m(k,\varphi)$ are Bloch functions and $q$ is the quasi-charge, which is analogous to the quasi-momentum in crystals. $\varepsilon_m(q)$ are the band energies that are periodic in $q$, and the period is the size of the first Brillouin zone $q\in [-0.5,0.5]$.

The presence of both the inductive term $\propto \varphi_\textnormal{J}^2$ and the driving $\propto \varphi_\textnormal{J}$ explicitly breaks the $2\pi$ translational symmetry of the junction Hamiltonian $\hat{H}_\textnormal{J}$, so that Bloch's theorem is not applicable to the total Hamiltonian $\hat{H}_2$, and we cannot work at fixed $q$. Nevertheless, the full continuum of quasi-charge states offers a convenient basis for calculations: not only it is able to easily handle the band structure of the spectrum, but it also gives a dynamical meaning to $q$, which is consistent with the dual picture.

In standard treatments of the dual Shapiro steps \cite{Likharev1985_th, Averin1990} one focuses on the transmon regime $E_\textnormal{J}\gg E_\textnormal{C}$, in which the first Bloch band has the approximate expression $\varepsilon_0(q)\simeq -U_0\cos(2\pi q)$, with the bandwidth \cite{Koch2007}
\begin{equation}\label{eq:transm-bandwidth}
     U_0=E_\textnormal{C} 2^5\sqrt{\frac{2}{\pi}}\left(\frac{E_\textnormal{J}}{2E_\textnormal{C}}\right)^{3/4}e^{-\sqrt{8E_\textnormal{J}/E_\textnormal{C}}}.
\end{equation}
In this regime, the bandwidth is much smaller than the separation from the first band $\varepsilon_1-\varepsilon_0\simeq \sqrt{8E_\textnormal{C}E_\textnormal{J}}$. This justifies the one-band approximation, where inter-band couplings are completely neglected. This approximation leaves in \eqref{ham_dual_flux} the Hamiltonian of the dual element of a Josephson junction: this means that phase and charge are interchanged and that the \textit{potential scale} $E_\textnormal{J}$ is replaced by $U_0$, while the \textit{mass scale} $4E_\textnormal{C}$ is replaced by the inductive energy $E_\textnormal{L}/2$.

It is worth emphasizing that the duality is approximate in two ways, since away from the transmon limit one can have both multi-band dynamics and a non-sinusoidal band. Given that recent experimental efforts focused on relatively small $E_\textnormal{J}/E_\textnormal{C}$ ratios \cite{Crescini2023,kaap2024demonstration}, it is important to include both of these effects in the theoretical treatment.

\subsection{Multi-band mean field dynamics}

To approach the dynamics of Hamiltonian \eqref{ham_dual_flux} without the single band approximation, we start from the full Heisenberg equations of motion. To take advantage of Bloch theorem, we employ the quasi-charge representation, analogous to the quasi-momentum representation in solid state physics \cite{Jones2003-zy,Blount1962,Koch2009}. We use the following splitting of the junction phase and charge operators:
\begin{equation}\label{qc_rep}
    \begin{cases}
        \hat{\varphi}_\textnormal{J}=\hat{\phi}+\hat{\Omega} \\
        \hat{n}_\textnormal{J}=\hat{q}+\hat{\mathcal{N}} \,,
    \end{cases}
\end{equation}
where $\hat\phi$ and $\hat q$ describe the intra-band dynamics, while $\hat\Omega$ and $\hat{\mathcal{N}}$ connect diferent bands. These operators are defined in terms of their matrix elements on Bloch functions, reported in Appendix \ref{appendixB}. In order to obtain a closed set of equations of motion, we need to consider the Heisenberg equations of the operators $\hat{a}_i$, $\hat{q}$, $\hat{\varphi}_\textnormal{J}$, and of the band-jump operators $\ket{m,q}\bra{m',q'}$. These are written in equation \eqref{eq:heisenberg_eq_dual}.

Analogously to the direct case, we want to first consider the mean field dynamics of these equations, while also taking into account the inter-band dynamics. To this end, we take the expectation values of the Heisenberg equations on a state of the following form:
\begin{equation}\label{dual_MF_ansatz}
    \ket{\psi}=\bigotimes_i\ket{\alpha_i}  \otimes\sum_m c_m \vert m,q\rangle\,.
\end{equation}
In other words, we treat the dynamics of the photonic component and of the quasi-charge in a classical fashion, while the full quantum mechanical inter-band dynamics is retained by assuming a generic superposition of Bloch states in different bands.

Such multi-band mean-field dynamics is described by the following set of differential equations:
\begin{widetext}
    \begin{equation}\label{eom_dual_flux}
        \begin{cases}
        \langle\dot{\hat{a}}_i\rangle=-i\tilde\omega_i\langle\hat{a}_i\rangle-\frac{i}{\hbar}\tilde g_i \left[\langle\hat{\varphi}_\textnormal{J}\rangle-\varphi_\textnormal{D}(t)\right]\\[10pt]
        \langle \dot{\hat{q}} \rangle=-\frac{E_\textnormal{L}}{\hbar}\langle \hat{\varphi}_\textnormal{J}\rangle+\frac{E_\textnormal{L}}{\hbar}\varphi_\textnormal{D}(t)-\frac{1}{\hbar}\sum_i \tilde g_i(\langle\hat{a}_i\rangle+\langle\hat{a}^\dagger_i\rangle)\\[10pt]
        \langle\dot{\hat{\varphi}}_\textnormal{J}\rangle=\frac{1}{\hbar}\sum_m  \partial_q \varepsilon_m(q) \rho_{m,m}+\frac{i}{\hbar}\sum_{m,m'}^{(m\neq m')} \Delta\varepsilon_{m,m'}(q)\Omega_{m,m'}(q)\rho_{m,m'}\\[10pt]
        \dot{\rho}_{m,m'}=\frac{i}{\hbar}\Delta\varepsilon_{m,m'}(q)\rho_{m,m'}+i\frac{E_\textnormal{L}}{2\hbar}\sum_{\mu}\left[ \varphi^2_{\mu,m}(q)\rho_{\mu,m'}-\varphi^2_{m',\mu}(q)\rho_{m,\mu}\right]+\\[10pt]
        \quad \quad \ \ \ +\frac{i}{\hbar}\left[\sum_i \tilde g_i(\langle\hat{a}_i\rangle+\langle\hat{a}^\dagger_i\rangle)-E_\textnormal{L}\varphi_\textnormal{D}(t)\right] \sum_{\mu}\left[ \varphi_{\mu,m}(q)\rho_{\mu,m'}-\varphi_{m',\mu}(q)\rho_{m,\mu}\right] \,.
    \end{cases}
    \end{equation}
\end{widetext}
These are written in terms of the $q$-dependent matrix elements of the previously introduced operators $\Omega_{m,m'}(q)$, $\varphi_{m,m'}(q)=\langle m,q | \hat{\varphi}_\textnormal{J} |m',q \rangle$, $\varphi^2_{m,m'}(q)=\langle m,q | \hat{\varphi}^2_\textnormal{J} |m',q \rangle$. We also defined $\Delta \varepsilon _{m,m'}(q)=\varepsilon_m(q)-\varepsilon_{m'}(q)$ and introduced the matrix elements of the reduced density matrix of the bands degrees of freedom $\rho_{m,n}=c^*_m c_n$. 

These equations greatly simplify to a form similar to the ones of the direct case \eqref{eom_shap_flux} if one makes the single-band approximation, i.e. if one neglects higher bands and imposes $\rho_{00}=\abs{c_0}^2=1$. Correspondingly, the equation for the expectation value of the flux reduces to just a single term representing the semi-classical group velocity in the first band, namely $\langle \dot{\hat{\varphi}}_\textnormal{J}\rangle=\frac{1}{\hbar}\partial_q \varepsilon_0(q)\approx \frac{2\pi U_0}{\hbar}\sin(2\pi q)$, where the last approximation holds in the transmon regime ($E_\textnormal{J} \gg E_\textnormal{C}$), in which the ideal duality is approached. In this regime the junction effectively behaves as a nonlinear capacitor \cite{Pechenezhskiy2020}.

In the single band approximation the junction charge $\hat n_\textnormal{J}$ and the quasi-charge $\hat q$ become indistinguishable, but more generally the expectation value of the current through the junction is given by:
\begin{equation}
    \langle\hat{I}_\textnormal{J}\rangle=2e \langle\dot{\hat{n}}_\textnormal{J}\rangle=2e\left(\langle\dot{\hat{q}}\rangle+\langle\dot{\hat{\mathcal{N}}}\rangle \right) \,,
\end{equation}
where the mean-field equation of motion for $\hat{\mathcal{N}}$ is
\begin{equation}    \langle\dot{\hat{\mathcal{N}}}\rangle=\sum_{m,m'}\mathcal{N}_{m,m'}(q)\dot{\rho}_{m,m'} \,.
\end{equation}
The approximation $\langle\hat{I}_\textnormal{J}\rangle \simeq 2e\langle\dot{\hat{q}}\rangle$ is accurate as long as the contribution given by the inter-band dynamics is negligible. For the cases we consider in the following, we always found this extra term to give very small corrections to the current, even though the presence of the multi-band dynamics can change the width of the plateaus.

\begin{figure*}[t!]
    \centering
    \includegraphics[width= 0.8\textwidth]{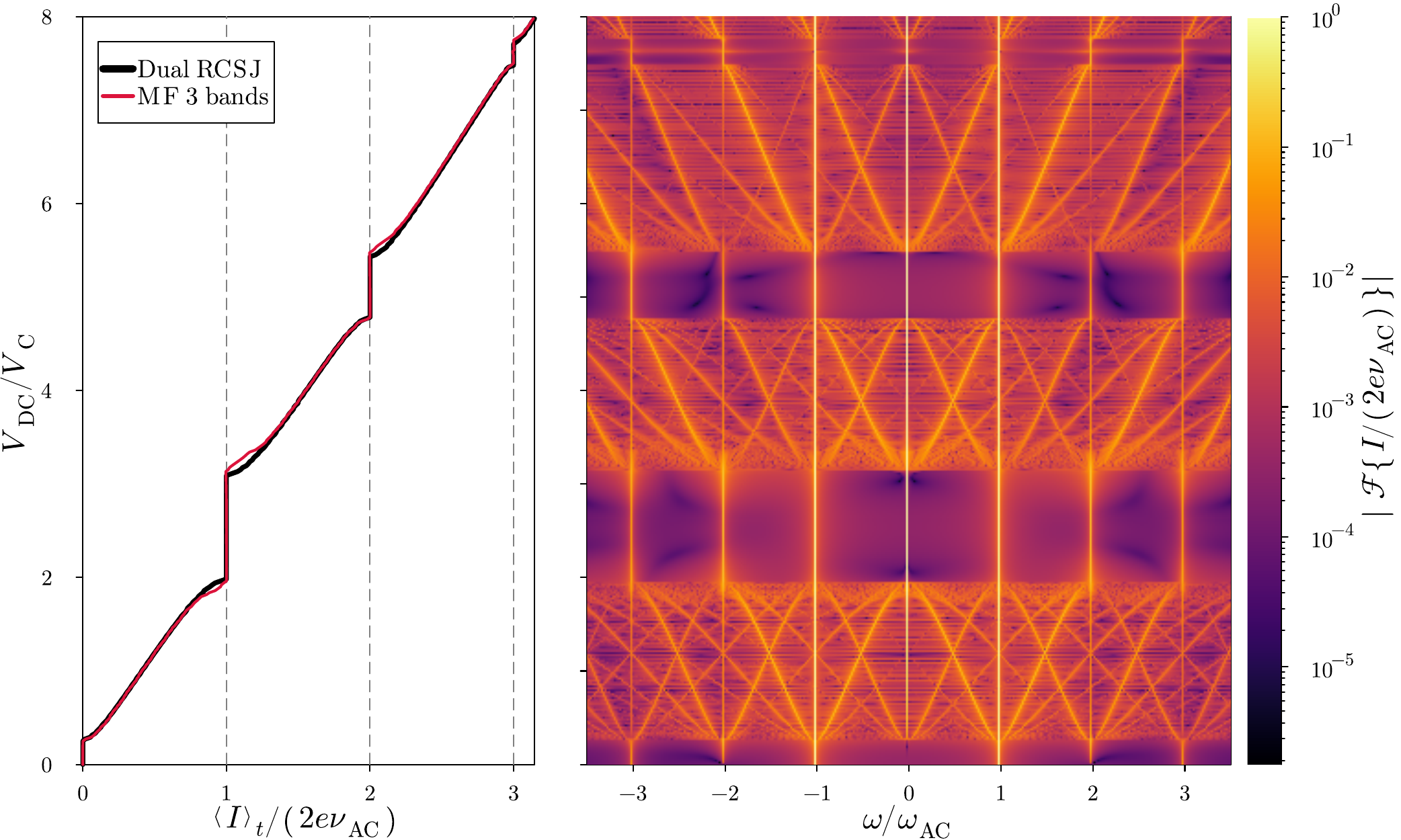}
    \caption{Left panel: dual Shapiro steps predicted by the multi-band mean-field dynamics for the junction coupled to a finite-size transmission line resonator and external inductance, as described by the system of equations in (\ref{eom_dual_flux}).
    The red curve depicts the solution of the dual RCSJ model \eqref{dual_diff_eq}, where $V_\textnormal{C}$ is the band-width of the lowest band and the resistance $R$ is equal to the impedance   $Z$ of the transmission line. Parameters: $E_\textnormal{C}/E_\textnormal{L}=400$, $E_\textnormal{J}/E_\textnormal{C}=10$, $Z/R_\textnormal{Q}=5$, $\hbar \omega_\textnormal{P}/E_\textnormal{L}=20\pi$, $\hbar\Delta/E_\textnormal{L}= 2\pi/75$, $V_\textnormal{AC}/V_\textnormal{C}=5$, $\hbar\omega_\textnormal{AC}/E_\textnormal{L}=2\pi$. The initial conditions for all variables of (\ref{eom_dual_flux}) are set to zero. The time averages are taken within this window over 50 periods of the AC source, excluding the first 15 to avoid transient effects. Unlike the direct case, where the mean-field dynamics and RCSJ model perfectly matches, in the dual case there are differences due to the non-negligible role of multiple bands. 
    Right panel: logarithmic scale color plot of the absolute value of the Fourier transform of the output current versus voltage and frequency.}
    \label{fig:dual_fourier}
\end{figure*}

\begin{figure}[h!]
    \centering
      \includegraphics[width=\columnwidth]{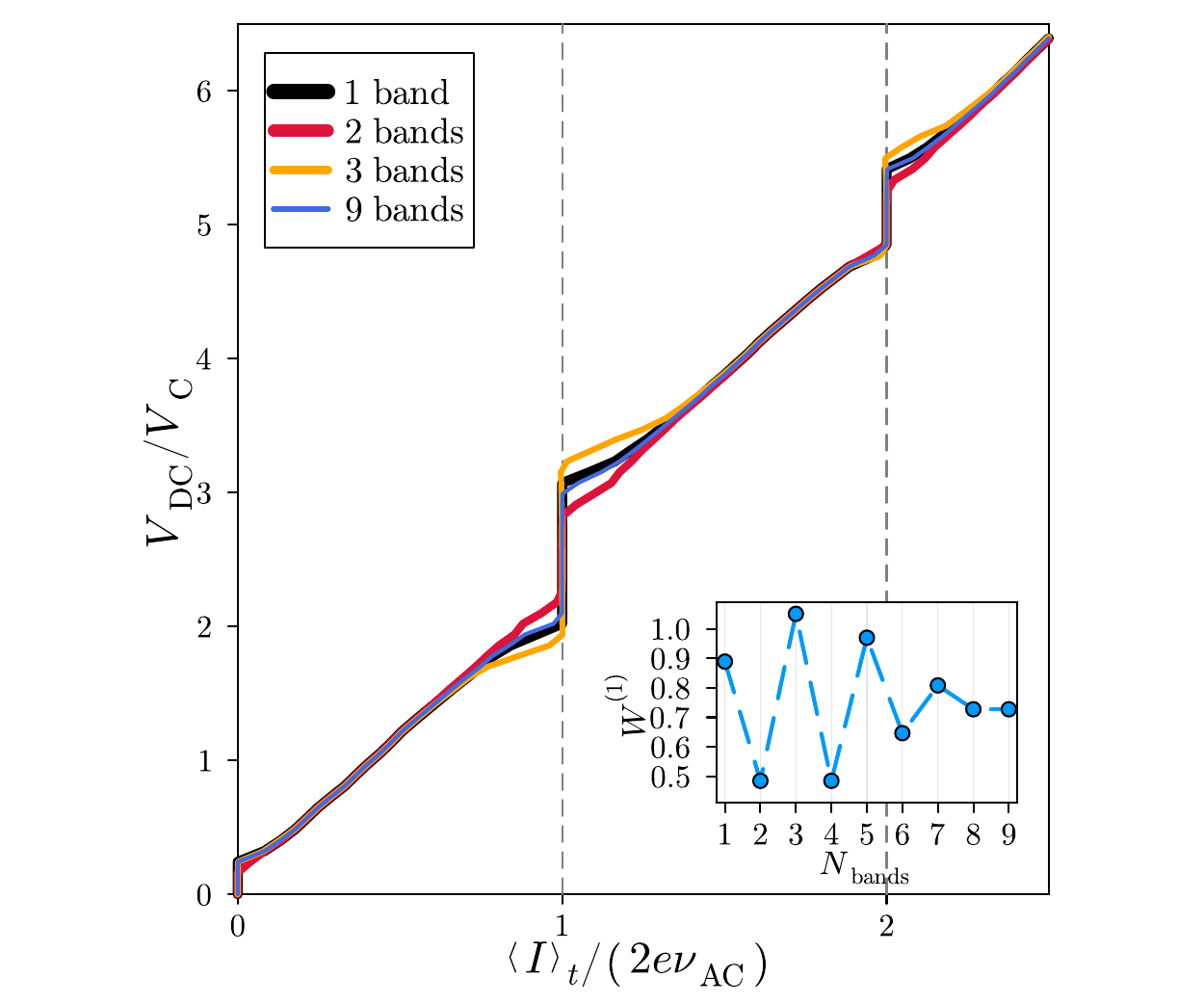}
    \caption{Comparison of dual Shapiro steps for different number of bands, namely 1, 2, 3, and 9. The inset, instead, shows the width of the first plateau as a function of the number of bands. We clearly see the asymmetry between odd and even bands, and convergence is reached for $N_\textnormal{bands}=8$. The time evolution is performed over 75 cycles of the AC driving, and the averages are performed over the last 50 cycles. We define a plateau as the set of points whose derivative is smaller than a given cutoff, i.e. 
    $\frac{V_\textnormal{C}}{2e \nu_\textnormal{AC}}\frac{d\langle I \rangle_t}{dV_\textnormal{DC}}<10^{-1}$. This measure of flatness selects only the central part of the plateau, and cuts out the edges, allowing us to consistently measure the width. Parameters: $E_\textnormal{C}/E_\textnormal{L}=400$, $E_\textnormal{J}/E_\textnormal{C}=10$, $Z/R_\textnormal{Q}=5$, $\hbar \omega_\textnormal{P}/E_\textnormal{L}=20\pi$, $\hbar\Delta/E_\textnormal{L}= 2\pi/75$, $V_\textnormal{AC}/V_\textnormal{C}=5$, $\hbar\omega_\textnormal{AC}/E_\textnormal{L}=2\pi$.}
    \label{fig:dual_band_dep}
\end{figure}

We solve the multi-band mean-field equations \eqref{eom_dual_flux} numerically, analogously to the direct case. The necessary energy bands and matrix elements are determined by numerically calculating the eigenstates of the Josephson Hamiltonian for different quasi-charges, namely $\hat{H}^q_\textnormal{J}=4 E_\textnormal{C} (\hat{n}_\textnormal{J}-q)^2-E_\textnormal{J} \cos(\hat\varphi_\textnormal{J})$, within the interval $\varphi_\textnormal{J} \in (-\pi,\pi]$. It is important to note that the matrix elements of operators that are not $2\pi$-periodic in $\varphi_\textnormal{J}$, such as $\varphi_{m,m'}(q)$ and $\varphi^2_{m,m'}(q)$, may depend on the total volume of the $\varphi_\textnormal{J}$ domain, that is in principle infinite (extended phase). One way of computing them is to consider a domain composed of several $2\pi$ cells, impose periodic boundary conditions on the whole volume and check their convergence as the number of cells is increased. Attention must be paid to the diagonal ones $\bra{m,q} \hat{\phi} \ket{m,q}$ and $\bra{m,q} \hat{\phi}^2 \ket{m,q}$. The first ones are zero because of symmetry, while the second ones are indeed volume-dependent. However, the equations of motion include only differences between these elements, which turn out to be volume-independent. More details about these matrix elements are given in Appendix \ref{appendixB}.


In the left panel of Fig.\ref{fig:dual_fourier}, we display the IV characteristic obtained by using the same averaging method as in the direct case. Here, the current dual Shapiro steps appear at multiples of the driving frequency. This is compared with the classical dual RCSJ model, showing that the plateaus occur in a very similar fashion even when the multi-band dynamics is taken into account.

With respect to the direct case, here we need to ensure convergence of the current-voltage curve with the number of bands, and a comparison of the results is shown in Fig.\ref{fig:dual_band_dep}. One can see that the number of bands essentially influences only the edges of the plateaus. Note that there is an odd/even asymmetry in the convergence process, that is slower for an even number of bands. This is visible in the fact that the two-bands result predicts significantly narrower plateaus. The number of bands needed for a good convergence also depends on the temporal window over which the output voltage is averaged. For short enough times the number of considered bands does not matter. For longer times such as the ones considered in Fig.\ref{fig:dual_band_dep}, instead, the first band population decreases and higher bands start to influence the dynamics. A larger number of bands is also needed for smaller values of $E_\textnormal{J}/E_\textnormal{C}$. However, also for $E_\textnormal{J}/E_\textnormal{C}\sim 5$ we found that the center of the plateaus is robust with respect to the multi-band dynamics and independent on the number of bands used for the computation.


In the right panel of Fig.\ref{fig:dual_fourier}, we show the absolute value of the Fourier transform in time of the current for each DC voltage driving amplitude. Similar to the direct case, the plateaus originate from synchronization domains where the only significant frequency components are the harmonics of the driving frequency. With respect to the corresponding plot for the direct case (top panel of Fig.\ref{fig:direct_shapiro_MF}), the boundaries of the synchronization intervals are less sharp, reflecting the approximate nature of the duality.


        The dynamics between bands is primarily governed by two ratios: $\mathrm{min}_\textnormal{Q}(\varepsilon_1(q)-\varepsilon_0(q))/E_\textnormal{L}$ and $E_\textnormal{J}/E_\textnormal{C}$. The former ratio determines how inductance influences the coupling between bands—the greater this ratio, the broader the steps, as numerically verified (not shown). Therefore, a large inductance is required so that $E_\textnormal{L}\ll E_\textnormal{J},E_\textnormal{C}$. On the other hand, $E_\textnormal{J}/E_\textnormal{C}$ affects the shape and spacing of the bands. A low value leads to strong inter-band coupling that disrupts the plateaus. Theoretically, the transmon limit where $E_\textnormal{J}/E_\textnormal{C}\gg 1$ provides optimal band separation. However, the width of the plateaus, proportional to the bandwidth $U_0$, becomes exponentially small with increasing $E_\textnormal{J}/E_\textnormal{C}$ (equation \eqref{eq:transm-bandwidth}). Our results show that moderate inter-band dynamics does not disrupt synchronization, so that an intermediate value of $E_\textnormal{J}/E_\textnormal{C}\sim 5-10$ can be used to achieve larger dual steps.

\section{Beyond mean-field: quantum and thermal fluctuations}\label{sec:fluctuations}
Up to this point, our analysis has focused  on mean-field predictions, incorporating fully quantum aspects only through the inter-band dynamics in the dual case. However, relying solely on a mean-field framework is not sufficient to predict the occurrence of robust steps, useful for metrological purposes. Indeed, quantum and thermal fluctuations are necessarily present, and may have a key role that needs to be accounted for. To include these in our theoretical analysis, we apply a stochastic approach based on the so-called truncated Wigner approximation (TWA), that was successfully applied in quantum optics \cite{gardiner2004quantum} and to bosonic cold-atoms systems out of equilibrium \cite{Steel1998,Sinatra2002,blakie2008dynamics}.

\subsection{Truncated Wigner approach}
In the Truncated Wigner Approximation framework, the time evolution of an observable for a quantum system is approximated with the ensemble average of mean-field trajectories sampled according to a probability distribution in phase space  \cite{Polkovnikov2010}. Here we perform a Gaussian approximation of the initial Wigner function of the full system (junction plus modes), and then evolve it under driving.

For a simpler phase-space representation of the dynamics we introduce creation and annihilation operators $\hat b_\textnormal{J}$ and $\hat b_\textnormal{J}^\dagger$ for the junction degrees of freedom, whose expressions are different in the direct and dual case, as specified in the following sections. Within the TWA, we can express the expectation value of an operator $\hat{O}(\underline{\hat{\boldsymbol{a}}},\underline{\hat{\boldsymbol{a}}}^\dagger)$ as:
\begin{equation}
    \langle \hat{O}(\underline{\hat{\boldsymbol{a}}},\underline{\hat{\boldsymbol{a}}}^\dagger,t) \rangle \simeq \int \int d\underline{\boldsymbol{\alpha}}_0 d\underline{\boldsymbol{\alpha}}^*_0\mathcal{W} (\underline{\boldsymbol{\alpha}}_0,\underline{\boldsymbol{\alpha}}^*_0)O_\textnormal{W}(\underline{\boldsymbol{\alpha}}(t),\underline{\boldsymbol{\alpha}}^*(t))
\end{equation}
where we have introduced the vector of operators $\underline{\hat{\boldsymbol{a}}}=\left[\hat{a}_1, \dots,\hat{a}_N,\hat{b}_\textnormal{J}\right]^T$, the corresponding mean-field variables $\underline{\boldsymbol{\alpha}}=\left[\langle \hat{a}_1\rangle, \dots,\langle \hat{a}_N\rangle,\langle \hat{b}_\textnormal{J}\rangle \right]^T$, and the Weyl symbol $O_\textnormal{W}$ of the operator. In this expression, $\underline{\boldsymbol{\alpha}}(t)$ indicate the mean-field time evolution, and the zero subscript denotes the initial conditions at $t=0$.

It is important to note that in the considered framework it is not necessary to evolve in time the Wigner distribution $\mathcal{W}$, that is evaluated only at the initial time $t=0$. Instead, one can compute the time evolution of the expectation values by performing a sampling of the initial conditions $\{\underline{\boldsymbol{\alpha}}_0,\underline{\boldsymbol{\alpha}}^*_0\}$, evolving via the mean-field equations of motion and performing the ensemble average over $N_\textnormal{ens}$ different trajectories:
\begin{equation}
\langle \hat{O}(\underline{\hat{\boldsymbol{a}}},\underline{\hat{\boldsymbol{a}}}^\dagger,t) \rangle
\simeq \frac{1}{N_\textnormal{ens}} \sum_{i=1}^{N_\textnormal{ens}} O_\textnormal{W}(\underline{\boldsymbol{\alpha}}_{i}(t),\underline{\boldsymbol{\alpha}}_i^*(t)).
\end{equation}
This stochastic solution is well defined and  describes accurately the dynamics if $\mathcal{W}$ is positive-definite, so that it can be interpreted as a probability distribution.

In contrast to applications for open systems, in which dissipation introduces a dynamical noise term in the mean-field equations \cite{carusotto2013quantum}, here the quantum fluctuations are entirely given by the initial conditions. It is hence important to have a good approximation of the ground state of the system, that is however not known analytically. We follow an approach similar in spirit to the one used for Bose-Einstein condensates \cite{Steel1998}, in which the initial Wigner function is constructed with a Gaussian Ansatz in terms of the independent Bogoliubov normal modes.

For both the direct and dual cases, we start with the system of nonlinear differential equations (\ref{eom_shap_flux}) or (\ref{eom_dual_flux}) in the mean-field approximation, that can be formally written as:
\begin{equation}
    \begin{bmatrix} \underline{\dot{\boldsymbol{\alpha}}}(t) \\ \underline{\dot{\boldsymbol{\alpha}}}^* (t)\end{bmatrix}=\begin{bmatrix}\underline{\mathbf{F}}(\underline{\boldsymbol{\alpha}}(t),\underline{\boldsymbol{\alpha}}^*(t)) \\ \underline{\mathbf{F}}^*(\underline{\boldsymbol{\alpha}}(t),\underline{\boldsymbol{\alpha}}^*(t))\end{bmatrix} \,.
\end{equation}
Note that in the absence of driving, a stationary solution of these mean-field equations is given by $\underline{\boldsymbol{\alpha}}_\textnormal{MF}(t)=\underline{\boldsymbol{\alpha}}_\textnormal{MF}^* (t)=\underline{\boldsymbol{0}}$. We consider small deviations from this solution:
\begin{equation}
    \begin{bmatrix} \underline{\boldsymbol{\alpha}}(t) \\ \underline{\boldsymbol{\alpha}}^* (t)\end{bmatrix}=\begin{bmatrix} \underline{\boldsymbol{\alpha}}_\textnormal{MF}(t) \\ \underline{\boldsymbol{\alpha}}_\textnormal{MF}^* (t)\end{bmatrix}+\begin{bmatrix} \delta\underline{\boldsymbol{\alpha}}(t) \\ \delta\underline{\boldsymbol{\alpha}}^* (t)\end{bmatrix} \,.
\end{equation}


The evolution of the fluctuation at linear order in $\delta\underline{{\boldsymbol{\alpha}}},\ \delta\underline{{\boldsymbol{\alpha}}}^*$ is given by the Jacobian $\underline{\underline{\mathbf{J}}}$ of the differential equations calculated at the mean-field stationary point (this corresponds to the Bogoliubov problem used for cold-atomic systems):
\begin{equation}
    \begin{bmatrix} \delta\underline{\dot{\boldsymbol{\alpha}}}(t) \\ \delta\underline{\dot{\boldsymbol{\alpha}}}^*(t) \end{bmatrix} \simeq \left. \underline{\underline{\mathbf{J}}} \right|_{\underline{\boldsymbol{\alpha}}_\textnormal{MF}, \underline{\boldsymbol{\alpha}}_\textnormal{MF}^*} \begin{bmatrix} \delta\underline{\boldsymbol{\alpha}}(t) \\ \delta\underline{\boldsymbol{\alpha}}^*(t) \end{bmatrix} \,,
\end{equation}
The explicit expressions of the Jacobian for both the direct and dual cases are reported in the Appendix \ref{appendixC}.


The eigenmodes of the Jacobian can be obtained by diagonalizing it with a change of basis:
\begin{equation}
    \begin{bmatrix}
    \underline{\dot{\boldsymbol{\gamma}}} \\ \underline{\dot{\boldsymbol{\gamma}}}^*
    \end{bmatrix}
    =
\underline{\underline{\mathbf{M}}}^{-1}\underline{\underline{\mathbf{J}}}\ \underline{\underline{\mathbf{M}}}
\begin{bmatrix}
\underline{\boldsymbol{\gamma}} \\ \underline{\boldsymbol{\gamma}}^*
\end{bmatrix}
=
    \begin{bmatrix}
    diag(\underline{\tilde{\boldsymbol{\omega}}}) & 0 \\ 0 & -diag(\underline{\tilde{\boldsymbol{\omega}}}) 
\end{bmatrix}
\begin{bmatrix}
\underline{\boldsymbol{\gamma}} \\ \underline{\boldsymbol{\gamma}}^*
\end{bmatrix} \,.
\end{equation}
At the operatorial level, this diagonalization is a Bogoliubov transformation. This gives us independent modes that mix the environmental modes and the junction degrees of freedom.

An approximation for the Wigner function of the undriven system is hence given by taking an equilibrium thermal Wigner distribution for each normal mode. For the $n$-th mode with frequency $\tilde{\omega}_n$ this is given by \cite{gardiner2004quantum}:
\begin{equation*}
    \mathcal{W}_{n,T}(\gamma_n, \gamma^*_n)=\frac{2}{\pi}\tanh\left( \frac{\hbar \tilde{\omega}_n}{2k_\textnormal{B} T}\right)e^{ -2 |\gamma_n|^2 \tanh\left( \frac{\hbar \tilde{\omega}_n}{2k_\textnormal{B} T}\right)} \,.
\end{equation*}
Hence, we can perform the sampling of the initial conditions for the normal modes and obtain the ones in the original variables with the change of basis
\begin{equation}
    \begin{bmatrix} \underline{\boldsymbol{\alpha}}_0 \\ \underline{\boldsymbol{\alpha}}^*_0\end{bmatrix}=\underline{\underline{\mathbf{M}}}\begin{bmatrix} \underline{\boldsymbol{\gamma}}(0) \\ \underline{\boldsymbol{\gamma}}^*(0)\end{bmatrix} \,.
\end{equation}

\begin{figure}[t!]
    \centering
      \includegraphics[width=\columnwidth]{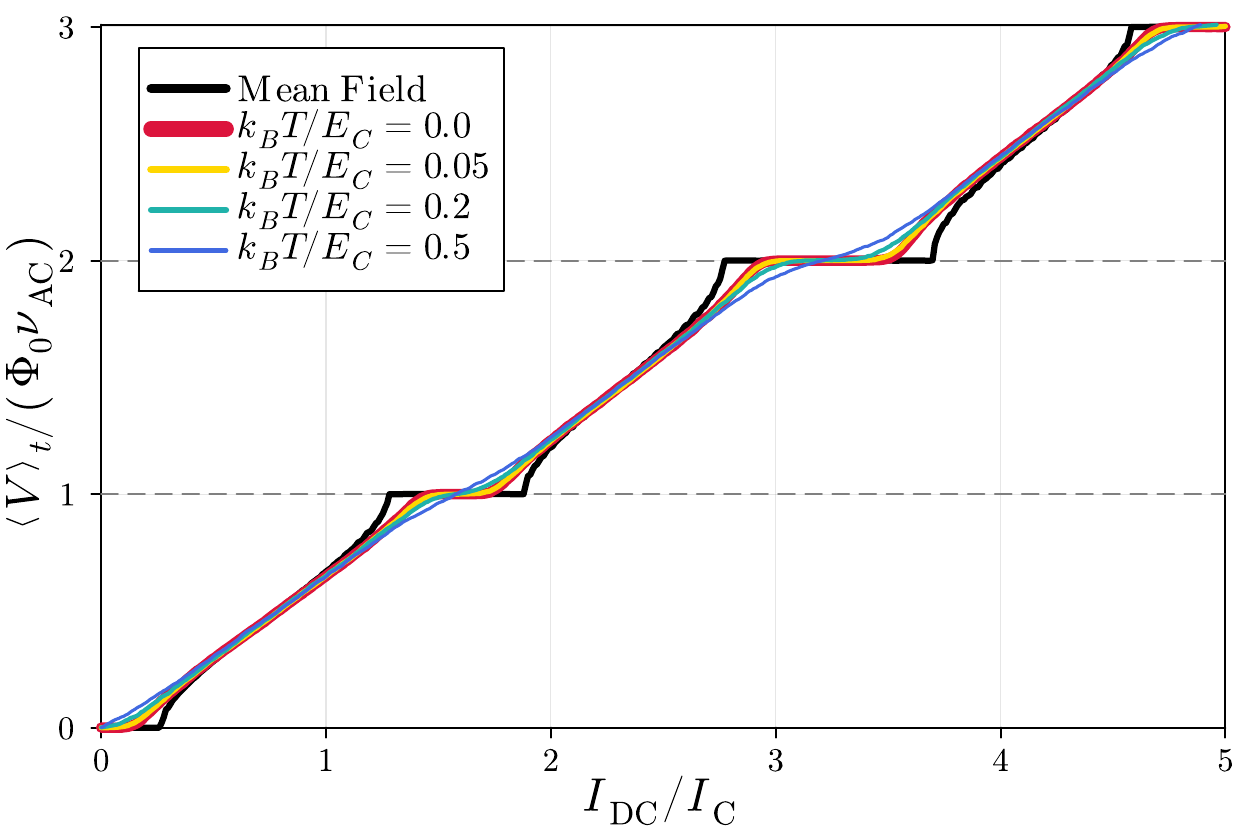}
    \caption{Direct Shapiro steps for a Josephson junction coupled to a low-impedance transmission line. The prediction of the mean-field theory is displayed by the thick black curve. The other curves include the effect of quantum and thermal fluctuations (see legend) calculated with the truncated Wigner approach. For an experimentally typical value of $E_\textnormal{C}/h=10 \ \textnormal{GHz}$, the temperatures correspond respectively to $0$, $25$, $100$, and $250 \ \textnormal{mK}$.
    Parameters: $E_\textnormal{J}/E_\textnormal{C}=1$, $Z/R_\textnormal{Q}=0.1$, $\hbar \omega_\textnormal{P}/E_\textnormal{C}=10$, $\hbar\Delta/E_\textnormal{C}=10^{-2}$, $I_\textnormal{AC}/I_\textnormal{C}=5$, $\hbar\omega_\textnormal{AC}/E_\textnormal{C}=1$.}
    \label{fig:shapiro_wigner}
\end{figure}

With this procedure we are constructing a Gaussian approximation of the initial state of the full system. A Gaussian Wigner function is  positive-definite, so the sampling procedure is well defined. Notice that we are making a harmonic approximation for the Josephson fluctuations, depending on the Josephson energy potential in the direct case and on the quasi-charge-dependent energy bands in the dual case. This method is expected to yield most accurate results in scenarios where either the flux or the charge is localized, that is in the correct regimes of $E_\textnormal{J}/E_\textnormal{C}$ and of impedance. We have checked numerically that, for the used parameters, our approximate vacuum and thermal states are stationary when the system is not driven (i.e. the means and variances of the resulting multi-mode Gaussian state do not change under free time evolution), as it must be within a consistent stochastic treatment.

Notice that this sampling procedure has the advantage of treating all degrees of freedom on the same footing, without giving a special treatment to junction variables. Moreover, in both the direct and dual cases the interaction term between the junction and the modes will introduce squeezing-inducing terms like $\hat{b}_\textnormal{J}a_i$ and $\hat{b}^\dagger_\textnormal{J}a_i^\dagger$; since these terms are quadratic, and the diagonalization of the Jacobian is equivalent to a Bogoliubov problem, as presented in Appendix \ref{appendixC}, the sampling procedure is also able to take into account multimode squeezing of the interacting system.

After the sampling of the initial conditions is performed, each of them is evolved with the mean field equations under driving. The IV characteristic is extracted analogously to the mean field case for each trajectory. The ensemble average of these curves is then performed to obtain the final result.


\subsection{Results for direct Shapiro steps}

\begin{figure}[t!]
    \centering
     \includegraphics[width=\columnwidth]{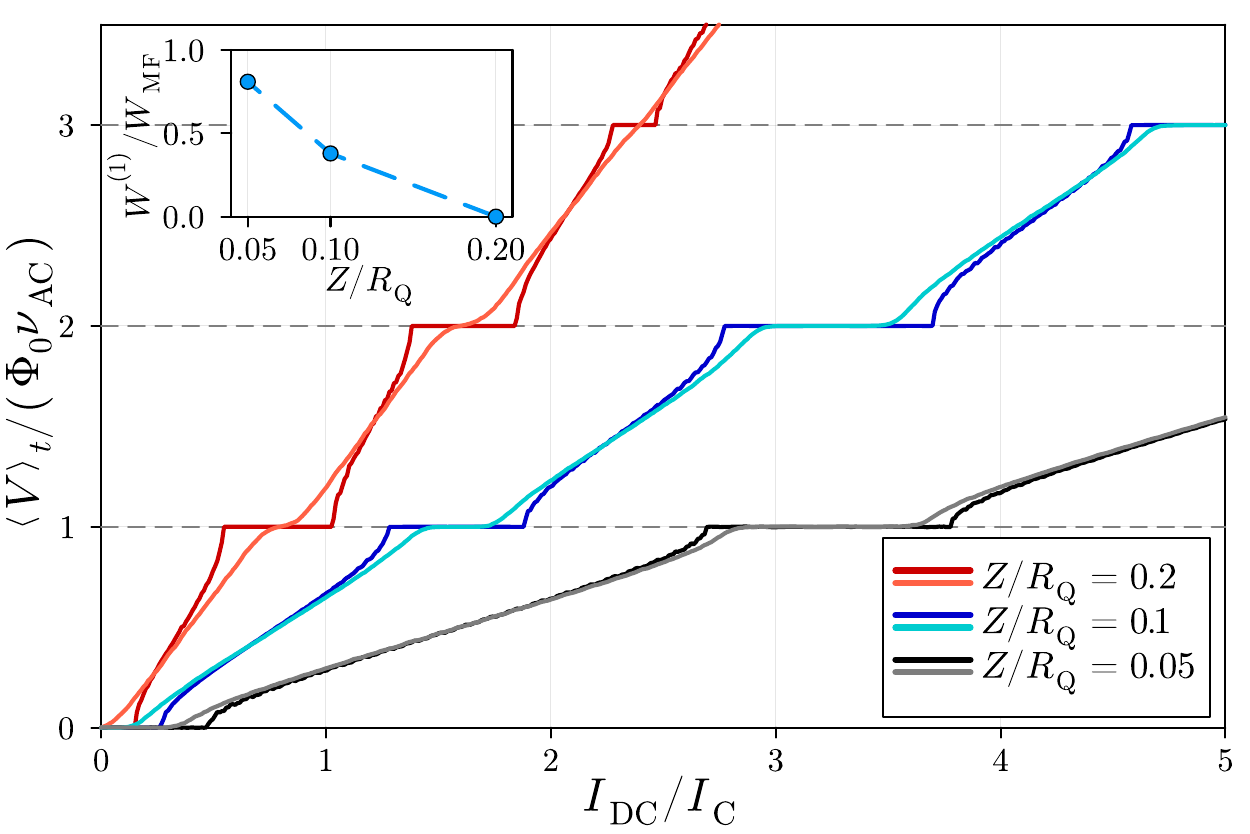}
    \caption{Dependence of direct Shapiro steps on the transmission line impedance. Darker curves show mean-field simulations at different impedances. Lighter shades, instead, correspond to simulations at zero temperature with the inclusion of vacuum fluctuations. The inset shows the width of the first plateau normalized by the corresponding mean field value as a function of impedance. The width is defined as the set of points such that $\frac{I_\textnormal{C}}{\Phi_0 \nu_\textnormal{AC}}\frac{d\langle V \rangle_t}{dI_\textnormal{DC}}<10^{-1}$. Other parameters are as in Fig. \ref{fig:shapiro_wigner}.}
    \label{fig:shapiro_wigner_Z}
\end{figure}

As an initial application, we employ the truncated Wigner method described in the previous section to analyze the direct case. The harmonic approximation is made in this case on the Josephson potential, so that we assume a localized phase. The corresponding creation and annihilaton operators for the junction degrees of freedom are introduced as $\hat{\varphi}_\textnormal{J}=\left( 2E_\textnormal{C}/E_\textnormal{J}\right)^{1/4}\left( \hat{b}_\textnormal{J}+\hat{b}_\textnormal{J}^\dagger\right)$ and $\hat{n}_\textnormal{J}=-\frac{i}{2}\left( E_\textnormal{J}/2E_\textnormal{C}\right)^{1/4}\left( \hat{b}_\textnormal{J}-\hat{b}_\textnormal{J}^\dagger\right)$.

In Fig.\ref{fig:shapiro_wigner} we compare the mean-field Shapiro steps with the results of the TWA approach, that incorporates vacuum fluctuations at zero temperature and thermal fluctuations at various temperatures. One can see that including vacuum fluctuations narrows the plateaus and smoothens the edges, although they remain clearly discernible. Including a finite temperature further shrinks these plateaus, and as the temperature is increased further the plateaus develop a slope and then vanish altogether. It is noteworthy that no parameter optimization was required to achieve these robust steps. In this analysis, we use a relatively small energy ratio, $E_\textnormal{J}/E_\textnormal{C}=1$. Employing higher values for $E_\textnormal{J}/E_\textnormal{C}$ might enhance the robustness. However, this robustness is not particularly surprising, given that the effect is already used as a metrological standard.
 
A crucial factor is instead the low impedance of the transmission line, which diminishes the phase fluctuations that disrupt synchronization. In Fig.\ref{fig:shapiro_wigner_Z} we compare the mean-field prediction with the zero-temperature result obtained with the truncated Wigner approximation for different impedances. One can see that the higher the impedance, the more the inclusion of zero-point fluctuations affects the plateaus. This confirms that a low impedance environment stabilizes the synchronization in the direct case.

\begin{figure}[t!]
    \centering
      \includegraphics[width=\columnwidth]{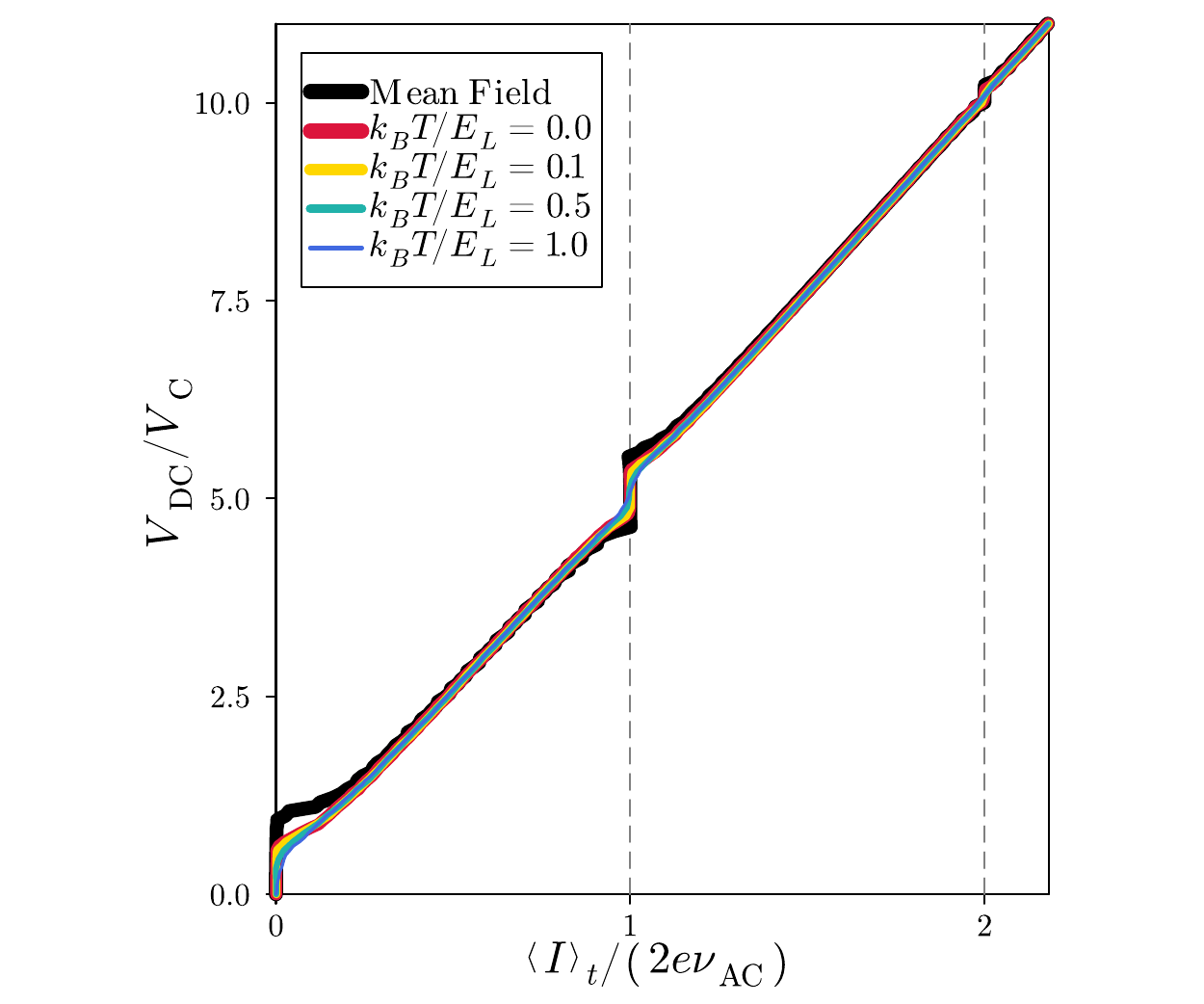}
    \caption{Dual Shapiro steps for a Josephson junction coupled to a finite-size high-impedance transmission line. The prediction of the mean-field theory is displayed by the thick black curve. The other curves include the effect of quantum and thermal fluctuations (see legend) calculated with the truncated Wigner approximation. For a value of $E_\textnormal{C}/h=400 \ \textnormal{GHz}$ (corresponding to a capacitance of roughly $0.05 \ \textnormal{fF}$), the temperatures correspond respectively to $0$, $4.8$, $24$, and $48 \ \textnormal{mK}$. Parameters: $E_\textnormal{C}/E_\textnormal{L}=400$, $E_\textnormal{J}/E_\textnormal{C}=10$, $Z/R_\textnormal{Q}=10$, $\hbar \omega_\textnormal{P}/E_\textnormal{L}=20\pi$, $\hbar\Delta/E_\textnormal{L}= 2\pi/50$, $V_\textnormal{AC}/V_\textnormal{C}=5$, $\hbar\omega_\textnormal{AC}/E_\textnormal{L}=2\pi$.}
    \label{fig:dual_shapiro_wigner}
\end{figure}

\subsection{Results for dual Shapiro steps}

\begin{figure}[t!]
    \centering
      \includegraphics[width= \columnwidth]{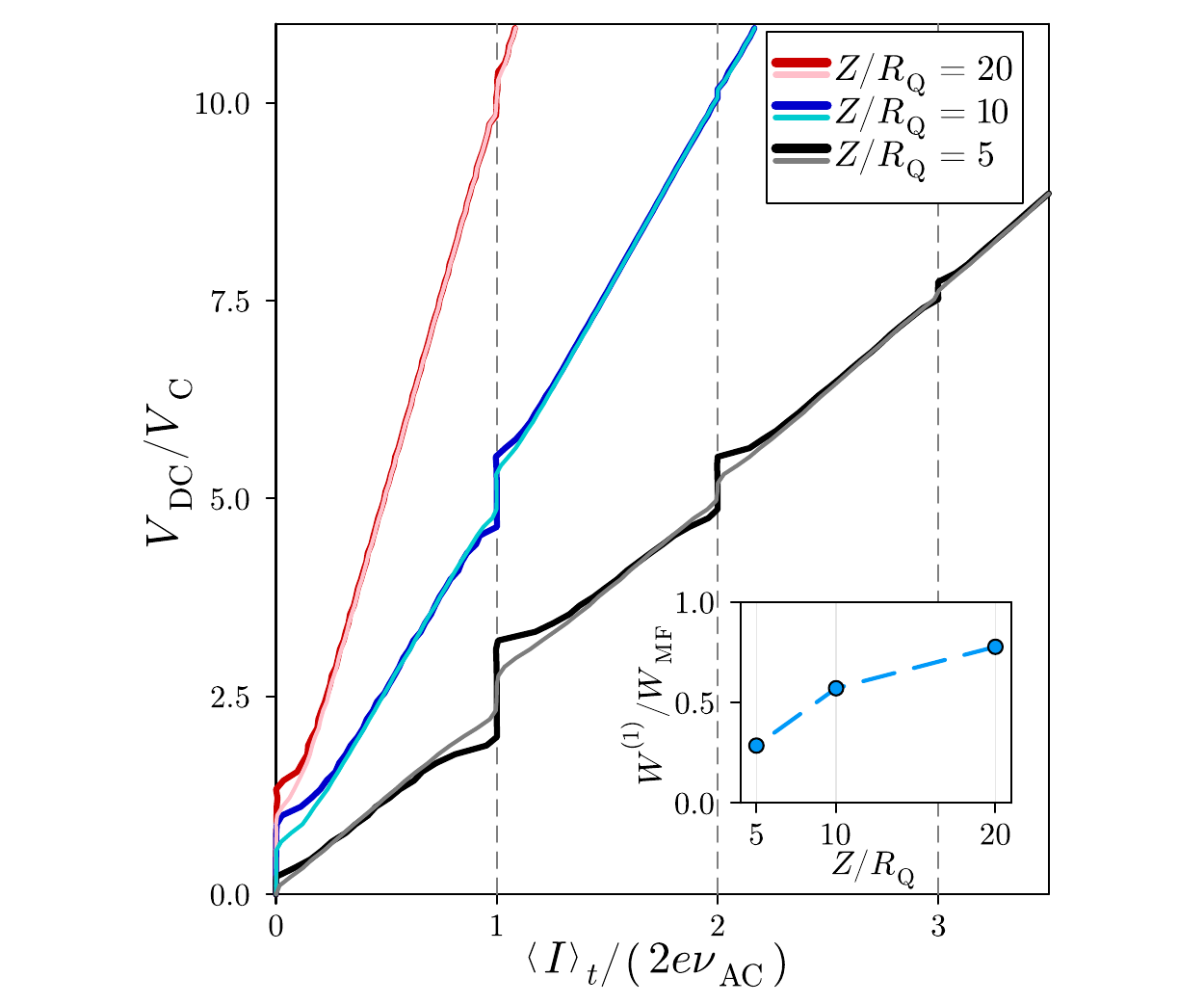}
    \caption{Dependence of dual Shapiro steps on the transmission line impedance. Darker curves show mean-field simulations at the three considered impedances. Lighter shades, instead, correspond to zero temperature with the inclusion of quantum vacuum fluctuations. The inset shows the width of the first plateau normalized by the corresponding mean field value as a function of impedance. The width is defined as the set of points such that $\frac{V_\textnormal{C}}{2e \nu_\textnormal{AC}}\frac{d\langle I \rangle_t}{dV_\textnormal{DC}}<10^{-1}$. Other parameters are as in Fig. \ref{fig:dual_shapiro_wigner}.}
    \label{fig:dual_shapiro_wigner_Z}
\end{figure}

The dual case presents more complexities, given by the multi-band nature of the problem. Quantum and thermal fluctuations can in fact disrupt synchronization both through charge delocalization and inter-band tunneling. Interestingly, even at high impedances $Z>R_\textnormal{Q}$, we observed that zero-point fluctuations can entirely eliminate the plateaus that are otherwise broad in the mean-field approximation.

In this case the harmonic approximation for the initial Wigner function is performed on the bands, so that we assume a localized quasicharge. The creation and annihilation operators are introduced as $\hat{\varphi}_\textnormal{J}=\tilde\eta\left( \hat{b}_\textnormal{J}+\hat{b}_\textnormal{J}^\dagger\right)$ and $\hat{q}=-\frac{i}{2}\tilde\eta^{-1}\left( \hat{b}_\textnormal{J}-\hat{b}_\textnormal{J}^\dagger\right)$, with $\tilde\eta=\left[\left(\sum_m \partial_q^2 \varepsilon_m(q=0) \rho_{m,m}\right)/4E_\textnormal{L}\right]^{1/4}$.

While at zero temperature the initial state Wigner function is analogous to the direct case, at finite temperatures one should also include a finite population in the higher bands. Apart from this important difference, the sampling of initial conditions is analogous to the direct case. The evolution under driving of these initial conditions is then performed with the mean-field multi-band equations \eqref{eom_dual_flux} and the IV characteristic is again obtained averaging over the single-trajectories ones.

In Fig.\ref{fig:dual_shapiro_wigner} we show a comparison between the mean-field multi-band equations and the stochastic simulations based on the truncated Wigner approximation at zero and finite temperature. The role of the quantum and thermal fluctuations is similar to the direct case, that is the curve at zero temperature shows a moderate reduction of the plateau compared to the mean-field case and an increase in temperature eventually leads to the destruction of the dual Shapiro step.

In the direct case the temperature was set with respect to the capacitive energy $E_\textnormal{C}$, while here the natural energy scale is the inductive one $E_\textnormal{L}$, with the plateaus disappearing for $k_\textnormal{B} T \sim E_\textnormal{L}$ for the considered impedance $Z/R_\textnormal{Q}=10$. Since we are choosing this energy scale to be small with respect to the charging energy, the absolute temperatures needed to have stable plateaus are much smaller than the ones in the direct case when comparing with the same typical values of $E_\textnormal{C}/h\sim 10\ 
\textnormal{GHz}$. It is hence beneficial to consider ultra-small junctions with a larger capacitive energy.

Similarly to the direct case, the impedance of the environmnent is fundamental to ensure the robustness of the dual steps. In Fig.\ref{fig:dual_shapiro_wigner_Z} we show a comparison of the multi-band mean-field prediction with the multi-band TWA
result at zero temperature. One can see that having a very high impedance is crucial. In fact, also resistances well above the resistance quantum such as $Z/R_\textnormal{Q}=5$ are not enough to sufficiently suppress charge fluctuations, even at zero temperature. This strong dependence on the impedance is not captured by the multi-band mean field description. Notice that the use of a higher impedance allows for smaller $E_\textnormal{C}$, since the steps still occur for larger values of $k_\textnormal{B}T/E_\textnormal{L}$.

An important result of our calculations is that, even though the single band approximation and the duality are less valid at smaller $E_\textnormal{J}/E_\textnormal{C}$, synchronization still occurs and is robust with respect to fluctuations at high enough impedance. This is a regime that is typically not considered in theoretical works relying on the duality, but has the advantage of larger plateaus.

\section{Conclusions}\label{sec:conclusions}

In this work, we have presented a comprehensive theoretical study on the occurrence of Shapiro steps and their dual counterpart in Josephson junctions coupled to finite-size transmission line resonators. While the direct steps are a well known and applied phenomenon, the observation and control of dual Shapiro steps are still a challenge.

We firstly developed a quantum treatment of direct Shapiro steps by employing a circuit QED approach. We derived the conditions under which these steps arise, confirming the results expected from the Resistively and Capacitively Shunted Junction (RCSJ) model. We then applied a similar treatment to dual Shapiro steps. Unlike previous studies, our model did not rely on the approximate charge-phase duality, but included full multi-band dynamics. Our results indicate that the multi-band dynamics, although complex, do not preclude the occurrence of dual Shapiro steps.

Furthermore, by incorporating quantum and thermal fluctuations through a truncated Wigner approach, we investigated how these fluctuations affect the stability and observability of Shapiro steps. Our results show that while direct Shapiro steps are robust against such fluctuations, dual Shapiro steps require optimized conditions, particularly concerning the impedance of the environment and the energy scale ratios within the Josephson junction.

Our framework is a valuable tool for the experimental efforts towards the metrological application of Josephson junctions in high-impedance environments. In fact, our approach allows to treat these systems without making use of the usual approximations, and is hence suitable for applications to new regimes and complex environments. For example, the current models can be expanded to include effects such as disorder, non-Markovian environmental interactions, and non-equilibrium dynamics.

In summary, our work not only provides theoretical tools for understanding the quantum dynamics of Josephson junctions coupled to complex environments, but also provides practical guidelines for their implementation in quantum technologies, specifically in the context of quantum metrology. 

\acknowledgements
We acknowledge support from the French project TRIANGLE (ANR-20-CE47-0011). 

{\it Note:} While completing this manuscript, two interesting arXiv preprints \cite{remez2024bloch,kurilovich2024quantum} appeared with independent works on the quantum theory of Bloch oscillations for resistively shunted junctions. Both these works focus on the transmon regime adopting a single-band approximation and taking the continuum limit of the environment.

\appendix

\section{Circuit QED Hamiltonians}\label{appendixA}

In this Appendix we discuss the derivation of the Hamiltonians describing the lumped-element circuits in Fig.\ref{fig:RCSJ} and Fig.\ref{fig:dual_circuit}. We use the circuit QED prescriptions to introduce the relevant variables \cite{devoret1995quantum,blais2021circuit}

\subsection{Circuit exhibiting direct Shapiro steps}
Here we derive Hamiltonian (\ref{ham_shap_flux}), describing the lumped-element circuit depicted in Fig.\ref{fig:RCSJ}, where we have defined the fluxes across the capacitors to be the independent degrees of freedom. The corresponding Lagrangian can be written in matrix form as:
\begin{align}
    \mathcal{L}&=\frac{1}{2}\frac{\hbar^2}{8E_\textnormal{C}}\boldsymbol{\dot{\varphi}}^T\mathbf{C}\boldsymbol{\dot{\varphi}}-\frac{1}{2}\frac{\hbar^2}{8E_\textnormal{C}}\boldsymbol{\varphi}^T\mathbf{L^{-1}}\boldsymbol{\varphi}+\frac{1}{2}\frac{\hbar^2}{8E_\textnormal{C}} \dot{\varphi}_\textnormal{J}^2+ \nonumber \\&-\frac{1}{2}E_\textnormal{L} \varphi^2_\textnormal{J} +E_\textnormal{J} \cos(\varphi_\textnormal{J})+E_\textnormal{L}\varphi_\textnormal{J} \varphi_1 +\frac{\hbar}{2e}I(t)\varphi_\textnormal{J} \,,
\end{align}
where $E_\textnormal{C}=\frac{e^2}{2C_\textnormal{J}}$, $E_\textnormal{L}=\left(\frac{\hbar}{2e}\right)^2\frac{1}{L}$, and $\boldsymbol{\varphi}=[\varphi_1,\varphi_2,\dots,\varphi_N]^T$ is a vector including the flux variables for the transmission line. The capacitance and inductance matrices read:
\begin{equation}
    \mathbf{C}=\frac{2E_\textnormal{C} C}{e^2} \boldsymbol{\mathds{1}}\,, \ 
\mathbf{L^{-1}}=\frac{2E_\textnormal{C}}{e^2L}\begin{bmatrix}
    2  & -1  &        &     \\ 
   -1  &  2  &   -1   &     \\ 
       & -1  & \ddots &     \\ 
       &     &        & 1 
 \end{bmatrix}\,.
\end{equation}
Notice that the pre-factors have been introduced to make both $\boldsymbol{\varphi}$ and $\mathbf{C}$ dimensionless. Given  the plasma frequency $\omega_\textnormal{P}$, the characteristic impedance $Z$ and the free spectral range $\Delta$ associated with the transmission line, the capacitance, inductance and the number of modes can be expressed as follows:
\begin{equation}
    \begin{cases}
        C=\frac{2}{Z \omega_\textnormal{P}}  \\
        L=\frac{2 Z}{\omega_\textnormal{P}} \\
        N=\frac{\pi \omega_\textnormal{P}}{2 \Delta} \,.
    \end{cases}
\end{equation}

The second step is to solve the generalized eigenvalue problem $\mathbf{L^{-1}}\mathbf{P} = \boldsymbol{\omega^2} \mathbf{C}\mathbf{P}$, which allows us to express the vector of fluxes in terms of normal modes as $\boldsymbol{\varphi}=\mathbf{P}\boldsymbol{\Phi}$. In terms of these normal coordinates, the Lagrangian can be expressed as:
\begin{align}
    \mathcal{L}&=\frac{1}{2}\frac{\hbar^2}{8E_\textnormal{C}}\sum_{i=1}^N \left( \dot{\Phi}_i^2 -\omega_i^2 \Phi_i^2 \right)+\frac{1}{2}\frac{\hbar^2}{8E_\textnormal{C}} \dot{\varphi}_\textnormal{J}^2-\frac{1}{2}E_\textnormal{L} \varphi^2_\textnormal{J}+\nonumber \\&+E_\textnormal{J} \cos(\varphi_\textnormal{J})+E_\textnormal{L} \varphi_\textnormal{J} \sum_{i=1}^N P_{1,i}\Phi_i +\frac{\hbar}{2e}I(t)\varphi_\textnormal{J} \, .
\end{align}

The third step consists in introducing the conjugated variables $n_\textnormal{J}=\frac{1}{\hbar}\frac{\partial\mathcal{L}}{\partial \dot{\varphi}_\textnormal{J}}$ and $n_i=\frac{1}{\hbar}\frac{\partial\mathcal{L}}{\partial \dot{\Phi}_i}$, and performing a Legendre transformation to obtain the corresponding Hamiltonian.  

The final step is a canonical quantization by imposing the commutation relations $[\hat{\Phi}_i,\hat{n}_j]=i\delta_{i,j}$ for $i,j\in\{1,2,\dots,N \}$, and $[\hat{\varphi}_\textnormal{J},\hat{n}_\textnormal{J}]=i$ (all other commutators vanish).  By introducing bosonic creation and annihilation operators for the modes such that $[\hat{a}_i,\hat{a}^\dagger_j]=\delta_{i,j}$, we can express the normal coordinates and associated momenta for the transmission line as:
\begin{equation}
    \begin{cases}
        \hat{\Phi}_i=\sqrt{\frac{4E_\textnormal{C}}{\hbar \omega_i}}(\hat{a}_i+\hat{a}^\dagger_i) \\
        \hat{n}_i=-i\sqrt{\frac{\hbar \omega_i}{16E_\textnormal{C}}}(\hat{a}_i-\hat{a}^\dagger_i) \,.
    \end{cases}
\end{equation}
Finally, we get the quantum Hamiltonian:
\begin{align}
    \hat{H}&=\sum_i \hbar \omega_i \hat{a}^\dagger_i \hat{a}_i+4E_\textnormal{C}\hat{n}^2_\textnormal{J}+\frac{1}{2}E_\textnormal{L} \hat{\varphi}^2_\textnormal{J} -E_\textnormal{J} \cos(\hat{\varphi}_\textnormal{J}) + \nonumber\\
    &-\hat{\varphi}_\textnormal{J} \sum_{i=1}^N g_i (\hat{a}_i+\hat{a}^\dagger_i) -\frac{\hbar}{2e}I(t)\varphi_\textnormal{J}\,,
\end{align}
where we have introduced the couplings $g_i=E_\textnormal{L}\sqrt{\frac{4E_\textnormal{C}}{\hbar \omega_i}}P_{1,i}$.

\subsection{Circuit exhibiting dual Shapiro steps}
Since the derivation is analogous to the one presented in the previous section, we will not repeat the detailed intermediate steps. The Lagrangian for the dual circuit considered in this work reads:
\begin{align}
    \mathcal{L}&=\frac{1}{2}\frac{\hbar^2}{8E_\textnormal{L}}\boldsymbol{\dot{\varphi}}^T\mathbf{C}\boldsymbol{\dot{\varphi}}-\frac{1}{2}\frac{\hbar^2}{8E_\textnormal{L}}\boldsymbol{\varphi}^T\mathbf{L^{-1}}\boldsymbol{\varphi}+\frac{1}{2}\frac{\hbar^2}{8E_\textnormal{C}} \dot{\varphi}_\textnormal{J}^2-\frac{1}{2}E_\textnormal{L} \varphi^2_\textnormal{J} + \nonumber \\& +E_\textnormal{J} \cos(\varphi_\textnormal{J})-E_\textnormal{L} \varphi_\textnormal{J} [\varphi_1-\varphi_\textnormal{D}(t)]+E_\textnormal{L} \varphi_\textnormal{D}(t) \varphi_1\,,
\end{align}
where in this case $E_\textnormal{L}=\left(\frac{\hbar}{2e}\right)^2\frac{1}{L_\textnormal{J}}$, and 
\begin{equation}
\mathbf{C}=\frac{2E_\textnormal{L} C}{e^2} \boldsymbol{\mathds{1}}\,, \ 
\mathbf{L^{-1}}=\frac{2E_\textnormal{L}}{e^2L}\begin{bmatrix}
    1+L/L_\textnormal{J}  & -1  &        &     \\ 
   -1  &  2  &   -1   &     \\ 
       & -1  & \ddots &     \\ 
       &     &        & 2 
 \end{bmatrix}\,.
\end{equation}

\section{Quasi-charge representation and equations of motion for the dual case}\label{appendixB}

Here,  we detail the explicit form of the operators (\ref{qc_rep}) expressed in the quasi-charge representation. Recalling the formal expansion $\hat{O}=\sum_{m,m'}\int dk dk' \ket{m,k} \bra{m,k}\hat{O}\ket{m',k'}\bra{m',k'}$, all we need to do is specify the expression of the matrix elements for each operator:
\begin{equation}
    \begin{cases}
\bra{m,k}\hat{\phi}\ket{m',k'}=i\,\delta_{m,m'}\partial_k\delta(k-k')\\
\bra{m,k}\hat{\Omega}\ket{m',k'}=\delta(k-k')\,\Omega_{m,m'}(k)\\
\bra{m,k}\hat{q}\ket{m',k'}=k\, \delta_{m,m'}\delta(k-k')\\
\bra{m,k}\hat{\mathcal{N}}\ket{m',k'}=\delta(k-k')\mathcal{N}_{m,m'}(k)\,,
    \end{cases}
\end{equation}
with:
\begin{equation}
    \begin{cases}
        \Omega_{m,m'}(k)=i\int_0^{2\pi}d\varphi u^*_m(k,\varphi) \partial_k u_{m'}(k,\varphi) \\[10pt]
        \mathcal{N}_{m,m'}(k)=-i\int_0^{2\pi}d\varphi u^*_m(k,\varphi)\partial_{\varphi}u_{m'}(k,\varphi)\,.
    \end{cases}
\end{equation}
Equipped with these definitions, we are able to express the equations of motion in this representation:

\begin{widetext}
\begin{equation}\label{eq:heisenberg_eq_dual}
        \begin{cases}
    \dot{\hat{a}}_i=-i\omega_i\hat{a}_i-\frac{i}{\hbar}g_i \left[\hat{\varphi}_\textnormal{J}-\varphi_\textnormal{D}(t)\right]\\[10pt]
    \dot{\hat{q}}=-\frac{E_\textnormal{L}}{\hbar}\hat{\varphi}_\textnormal{J}+\frac{E_\textnormal{L}}{\hbar}\varphi_\textnormal{D}(t)-\frac{1}{\hbar}\sum_i g_i[\hat{a}_i+\hat{a}^\dagger_i]\\[10pt]
    \dot{\hat{\varphi}}_\textnormal{J}=\frac{i}{\hbar}\sum_{m,m'}\int dkdk' \langle m,k|\left[ \hat{H}_{J},\hat{\varphi}_\textnormal{J}\right]|m',k'\rangle |m,k\rangle \langle m',k'|\\[10pt]
\partial_t(|m,k\rangle \langle m',k'|)=\frac{i}{\hbar}\sum_{\mu,\mu'}\int d\kappa d\kappa' \langle \mu,\kappa|\left[ \hat{H},|m,k\rangle \langle m',k'| \right]|\mu',\kappa'\rangle |\mu,\kappa\rangle \langle \mu',\kappa'|\,,
    \end{cases}
\end{equation}
where the matrix elements of the commutators are given by:
\begin{align}
\begin{cases}
    \bra{m,k} \left[ \hat{H}_{J},\hat{\varphi}_\textnormal{J}\right]\ket{m',k'} =\Delta \varepsilon_{m,m'}(k,k')\left(\bra{m,k}\hat{\phi}\ket{m',k'}+\bra{m,k}\hat{\Omega}\ket{m',k'}\right)\\[10pt]
    \bra{\mu,\kappa} \left[ \hat{H},\ket{m,k}\bra{m',k'} \right]|\ket{\mu',\kappa'} = \delta_{m,\mu}\delta_{m',\mu'} \delta(k-\kappa)\delta(k'-\kappa') \Delta\varepsilon_{m,m'}(k,k') +
\\[10pt]\quad\quad+\frac{E_\textnormal{L}}{2}\left[ \langle \mu, \kappa|\hat{\varphi}^2_\textnormal{J}|m,k\rangle \delta_{m',\mu'}\delta(k'-\kappa')-\langle m',k'|\hat{\varphi}^2_\textnormal{J}|\mu',\kappa'\rangle \delta_{m,\mu}\delta(k-\kappa)\right] +\\[10pt]\quad\quad
+\left[\sum_i g_i (\hat{a}_i+\hat{a}^\dagger_i)-E_\textnormal{L}\varphi_\textnormal{D}(t)\right]\left[ \langle \mu, \kappa|\hat{\varphi}_\textnormal{J}|m,k\rangle \delta_{m',\mu'}\delta(k'-\kappa')-\langle m',k'|\hat{\varphi}_\textnormal{J}|\mu',\kappa'\rangle \delta_{m,\mu}\delta(k-\kappa)\right]\,,
\end{cases}
\end{align}
with $\Delta\varepsilon_{m,m'}(k,k')=\varepsilon_m(k)-\varepsilon_{m'}(k')$.
By taking the expectation value over a generic state $\ket{\psi}$ of the equations above, we obtain the following system:
\begin{equation}\label{eom_qc_rep}
    \begin{cases}
    \langle\dot{\hat{a}}_i\rangle=-i\omega_i\langle\hat{a}_i\rangle-\frac{i}{\hbar}g_i \left[\langle\hat{\varphi}_\textnormal{J}\rangle-\varphi_\textnormal{D}(t)\right]\\[10pt]
    \langle \dot{\hat{q}} \rangle=-\frac{E_\textnormal{L}}{\hbar}\langle \hat{\varphi}_\textnormal{J}\rangle+\frac{E_\textnormal{L}}{\hbar}\varphi_\textnormal{D}(t)-\frac{1}{\hbar}\sum_i g_i[\langle\hat{a}_i\rangle+\langle\hat{a}^\dagger_i\rangle] \\[10pt]
    \langle\dot{\hat{\varphi}}_\textnormal{J}\rangle=\frac{1}{\hbar}\sum_m  \int dk\partial_k \varepsilon_m(k) \rho_{m,m}(k,k)+\frac{i}{\hbar}\sum_{m,m'}^{(m\neq m')} \int dk\Delta\varepsilon_{m,m'}(k)\Omega_{m,m'}(k)\rho_{m,m'}(k,k)\\[10pt]
\dot{\rho}_{m,m'}(k,k')=\frac{i}{\hbar}\Delta\varepsilon_{m,m'}(k,k')\rho_{m,m'}(k,k')+i\frac{E_\textnormal{L}}{2\hbar}\sum_{\mu}\int d\kappa\left[ \langle\mu, \kappa|\hat{\varphi}^2_\textnormal{J}|m,k\rangle \rho_{\mu,m'}(\kappa,k')-\langle m',k'|\hat{\varphi}^2_\textnormal{J}|\mu, \kappa\rangle  \rho_{m,\mu}(k,\kappa)\right]+\\[10pt]
\quad \quad \quad+\frac{i}{\hbar}\left[\sum_i g_i (\langle\hat{a}_i\rangle+\langle\hat{a}^\dagger_i\rangle)-E_\textnormal{L}\varphi_\textnormal{D}(t)\right]\sum_{\mu}\int d\kappa\left[ \langle\mu, \kappa|\hat{\varphi}_\textnormal{J}|m,k\rangle \rho_{\mu,m'}(\kappa,k')-\langle m',k'|\hat{\varphi}_\textnormal{J}|\mu, \kappa\rangle \rho_{m,\mu}(k,\kappa)\right]
\end{cases}
\end{equation}
with $\rho_{m,m'}(k,k')=\braket{\psi | m,k}\braket{m',k' | \psi}=\psi_m^*(k)\psi_{m'}(k')$ the reduced density matrix for the matter sector. According to ansatz (\ref{dual_MF_ansatz}), the density matrix can be decomposed as $\rho_{m,m'}(k,k')=c_m^* c_{m'} \delta(k-q)\delta(k'-q)$, and substituting this form in the equations above, we obtain exactly the system of differential equations (\ref{eom_dual_flux}).

\end{widetext}

\section{Details about the Wigner trajectories}\label{appendixC}
\setcounter{equation}{0}
\renewcommand{\theequation}{C.\arabic{equation}}
In this appendix we report the explicit expressions of the Jacobian matrix for both the direct and the dual case. Introducing the notation:
\begin{equation*}
\underline{\boldsymbol{\beta}}=
    \begin{bmatrix} \underline{\boldsymbol{\alpha}} \\ \underline{\boldsymbol{\alpha}}^*\end{bmatrix}, \quad \underline{\mathbf{G}}(\underline{\boldsymbol{\beta}})=\begin{bmatrix}\underline{\mathbf{F}}(\underline{\boldsymbol{\beta}}) \\ \underline{\mathbf{F}}^*(\underline{\boldsymbol{\beta}})\end{bmatrix}\,,
\end{equation*}
the entries of the Jacobian matrix are defined as:
\begin{equation}
[\underline{\underline{\mathbf{J}}}]_{i,j}=\frac{\partial G_i}{\partial \beta_\textnormal{J}}\,.
\end{equation}
Hence, for the direct case we obtain:
\begin{equation}
    \underline{\underline{\mathbf{J}}}=\begin{bmatrix}
\underline{\underline{\mathbf{A}}}  &  \underline{\underline{\mathbf{B}}}     \\   -\underline{\underline{\mathbf{B}}} & -\underline{\underline{\mathbf{A}}}
 \end{bmatrix} \,,
\end{equation}
where:
\begin{equation}
    \underline{\underline{\mathbf{A}}}=-\frac{i}{\hbar}\begin{bmatrix}
\hbar\omega_1  &        &               &-\tilde{g}_1 \\ 
               & \ddots &               & \vdots       \\ 
               &        & \hbar\omega_N &-\tilde{g}_N \\ 
-\tilde{g}_1   & \dots  &-\tilde{g}_N   & (e_1) 
 \end{bmatrix} \,,
\end{equation}
and:
\begin{equation}
    \underline{\underline{\mathbf{B}}}=-\frac{i}{\hbar}\begin{bmatrix}
               &        &               &-\tilde{g}_1 \\ 
               &        &               & \vdots      \\ 
               &        &               &-\tilde{g}_N \\ 
-\tilde{g}_1   & \dots  &-\tilde{g}_N   & (e_2) 
 \end{bmatrix} \,,
\end{equation}
with $\tilde{g}_i=\eta g_i$, $\eta=\left(\frac{2E_\textnormal{C}}{E_\textnormal{J}}\right)^{1/4}$and  $\omega_\textnormal{J}=\frac{\sqrt{8 E_\textnormal{C} E_\textnormal{J}}}{\hbar}$ . Moreover, we have:
\begin{equation*}
\begin{cases}
    (e_1)=\eta^2 E_\textnormal{L} + \frac{\hbar \omega_\textnormal{J}}{2} +\eta^2 E_\textnormal{J} \cos\left[\eta\left(\langle \hat{b}_\textnormal{J} \rangle + \langle \hat{b}^\dagger_\textnormal{J} \rangle \right)\right] \\[10pt]
    (e_2)=\eta^2 E_\textnormal{L} - \frac{\hbar \omega_\textnormal{J}}{2} +\eta^2 E_\textnormal{J} \cos\left[\eta\left(\langle \hat{b}_\textnormal{J} \rangle + \langle \hat{b}^\dagger_\textnormal{J} \rangle \right)\right] \,.
\end{cases} 
\end{equation*}
In the case we are interested in, we need to evaluate the Jacobian along the mean-field trajectories at the initial time, in other words for vanishing initial conditions. Hence, the expressions above reduce to:
\begin{equation*}
\begin{cases}
    (e_1)=\eta^2 E_\textnormal{L} + \hbar \omega_\textnormal{J} \\[10pt]
    (e_2)=\eta^2 E_\textnormal{L}  \,.
\end{cases} 
\end{equation*}

As for the dual case, the matrices composing the Jacobian get modified in the following manner:
\begin{equation}
    \underline{\underline{\mathbf{A}}}'=-\frac{i}{\hbar}\begin{bmatrix}
\hbar\omega_1  &        &               &\tilde{g}_1' \\ 
               & \ddots &               & \vdots      \\ 
               &        & \hbar\omega_N &\tilde{g}_N' \\ 
\tilde{g}_1'   & \dots  &\tilde{g}_N'   & (e'_1) 
 \end{bmatrix} \,,
\end{equation}
and:
\begin{equation}
    \underline{\underline{\mathbf{B}}}'=-\frac{i}{\hbar}\begin{bmatrix}
               &        &               &\tilde{g}_1' \\ 
               &        &               & \vdots      \\ 
               &        &               &\tilde{g}_N' \\ 
\tilde{g}_1'   & \dots  &\tilde{g}_N'   & (e'_2) 
 \end{bmatrix} \,, \\[10pt]
\end{equation}
with $\tilde{g}'_i=\eta' g_i$, $\eta'=\left[ \frac{\left. \sum_m \partial_q^2 \varepsilon_m(q) \right|_{q=0} \rho_{m,m}}{4E_\textnormal{L}} \right]^{1/4}$, and:
\begin{align*}
\begin{cases}
    (e'_1) =&\eta'^2 E_\textnormal{L} + \frac{ \sum_m \partial_q^2 \varepsilon_m(q) \rho_{m,m}}{4\eta'^2}+ \\ &+\frac{i}{4\eta'^2} \sum_{m,m'}^{(m\neq m')} \partial_q [\Delta \varepsilon_{m,m'}(q)\Omega_{m,m'}(q)]\rho_{m,m'}\\[10pt]
    (e'_2) =&\eta'^2 E_\textnormal{L} - \frac{ \sum_m \partial_q^2 \varepsilon_m(q) \rho_{m,m}}{4\eta'^2}+ \\ &-\frac{i}{4\eta'^2} \sum_{m,m'}^{(m\neq m')} \partial_q [\Delta \varepsilon_{m,m'}(q)\Omega_{m,m'}(q)]\rho_{m,m'}  \,.
\end{cases} 
\end{align*}
In the case we are interested in, we need to evaluate the Jacobian along the mean-field trajectories at the initial time. Since the only time dependence is implicitly carried by $q(t)$, we need to evaluate all expressions above at $q(0)=0$. As for the density matrix, we assume a pure thermal state, so that the diagonal components are given by the usual Boltzmann weight $\rho_{m,m}=\frac{\exp \left[\frac{\varepsilon_m(0)}{k_B T}\right]}{\sum_m \exp\left[\frac{\varepsilon_m(0)}{k_B T}\right]}$, while the off-diagonal elements $\rho_{m,m'}$, with $m\neq m'$, are assumed to vanish.

\bibliography{theory_steps_v4.bib}

\begin{thebibliography}{55}%
\makeatletter
\providecommand \@ifxundefined [1]{%
 \@ifx{#1\undefined}
}%
\providecommand \@ifnum [1]{%
 \ifnum #1\expandafter \@firstoftwo
 \else \expandafter \@secondoftwo
 \fi
}%
\providecommand \@ifx [1]{%
 \ifx #1\expandafter \@firstoftwo
 \else \expandafter \@secondoftwo
 \fi
}%
\providecommand \natexlab [1]{#1}%
\providecommand \enquote  [1]{``#1''}%
\providecommand \bibnamefont  [1]{#1}%
\providecommand \bibfnamefont [1]{#1}%
\providecommand \citenamefont [1]{#1}%
\providecommand \href@noop [0]{\@secondoftwo}%
\providecommand \href [0]{\begingroup \@sanitize@url \@href}%
\providecommand \@href[1]{\@@startlink{#1}\@@href}%
\providecommand \@@href[1]{\endgroup#1\@@endlink}%
\providecommand \@sanitize@url [0]{\catcode `\\12\catcode `\$12\catcode
  `\&12\catcode `\#12\catcode `\^12\catcode `\_12\catcode `\%12\relax}%
\providecommand \@@startlink[1]{}%
\providecommand \@@endlink[0]{}%
\providecommand \url  [0]{\begingroup\@sanitize@url \@url }%
\providecommand \@url [1]{\endgroup\@href {#1}{\urlprefix }}%
\providecommand \urlprefix  [0]{URL }%
\providecommand \Eprint [0]{\href }%
\providecommand \doibase [0]{https://doi.org/}%
\providecommand \selectlanguage [0]{\@gobble}%
\providecommand \bibinfo  [0]{\@secondoftwo}%
\providecommand \bibfield  [0]{\@secondoftwo}%
\providecommand \translation [1]{[#1]}%
\providecommand \BibitemOpen [0]{}%
\providecommand \bibitemStop [0]{}%
\providecommand \bibitemNoStop [0]{.\EOS\space}%
\providecommand \EOS [0]{\spacefactor3000\relax}%
\providecommand \BibitemShut  [1]{\csname bibitem#1\endcsname}%
\let\auto@bib@innerbib\@empty
\bibitem [{\citenamefont {Likharev}\ and\ \citenamefont
  {Zorin}(1985)}]{Likharev1985_th}%
  \BibitemOpen
  \bibfield  {author} {\bibinfo {author} {\bibfnamefont {K.~K.}\ \bibnamefont
  {Likharev}}\ and\ \bibinfo {author} {\bibfnamefont {A.~B.}\ \bibnamefont
  {Zorin}},\ }\bibfield  {title} {\bibinfo {title} {Theory of the
  {{B}}loch-wave oscillations in small {{J}}osephson junctions},\ }\href
  {https://doi.org/10.1007/bf00683782} {\bibfield  {journal} {\bibinfo
  {journal} {Journal of Low Temperature Physics}\ }\textbf {\bibinfo {volume}
  {59}},\ \bibinfo {pages} {347–382} (\bibinfo {year} {1985})}\BibitemShut
  {NoStop}%
\bibitem [{\citenamefont {{{J}}osephson}(1962)}]{Josephson1962}%
  \BibitemOpen
  \bibfield  {author} {\bibinfo {author} {\bibfnamefont {B.}~\bibnamefont
  {{{J}}osephson}},\ }\bibfield  {title} {\bibinfo {title} {Possible new
  effects in superconductive tunnelling},\ }\href
  {https://doi.org/10.1016/0031-9163(62)91369-0} {\bibfield  {journal}
  {\bibinfo  {journal} {Physics Letters}\ }\textbf {\bibinfo {volume} {1}},\
  \bibinfo {pages} {251–253} (\bibinfo {year} {1962})}\BibitemShut {NoStop}%
\bibitem [{\citenamefont {{{S}}hapiro}(1963)}]{Shapiro1963}%
  \BibitemOpen
  \bibfield  {author} {\bibinfo {author} {\bibfnamefont {S.}~\bibnamefont
  {{{S}}hapiro}},\ }\bibfield  {title} {\bibinfo {title} {{{J}}osephson
  currents in superconducting tunneling: The effect of microwaves and other
  observations},\ }\href {https://doi.org/10.1103/PhysRevLett.11.80} {\bibfield
   {journal} {\bibinfo  {journal} {Phys. Rev. Lett.}\ }\textbf {\bibinfo
  {volume} {11}},\ \bibinfo {pages} {80} (\bibinfo {year} {1963})}\BibitemShut
  {NoStop}%
\bibitem [{\citenamefont {Grimes}\ and\ \citenamefont
  {{{S}}hapiro}(1968)}]{Grimes1968}%
  \BibitemOpen
  \bibfield  {author} {\bibinfo {author} {\bibfnamefont {C.~C.}\ \bibnamefont
  {Grimes}}\ and\ \bibinfo {author} {\bibfnamefont {S.}~\bibnamefont
  {{{S}}hapiro}},\ }\bibfield  {title} {\bibinfo {title} {Millimeter-wave
  mixing with {{J}}osephson junctions},\ }\href
  {https://doi.org/10.1103/physrev.169.397} {\bibfield  {journal} {\bibinfo
  {journal} {Physical Review}\ }\textbf {\bibinfo {volume} {169}},\ \bibinfo
  {pages} {397–406} (\bibinfo {year} {1968})}\BibitemShut {NoStop}%
\bibitem [{\citenamefont {Pikovsky}\ \emph {et~al.}(2001)\citenamefont
  {Pikovsky}, \citenamefont {Rosenblum},\ and\ \citenamefont
  {Kurths}}]{Pikovsky2001}%
  \BibitemOpen
  \bibfield  {author} {\bibinfo {author} {\bibfnamefont {A.}~\bibnamefont
  {Pikovsky}}, \bibinfo {author} {\bibfnamefont {M.}~\bibnamefont
  {Rosenblum}},\ and\ \bibinfo {author} {\bibfnamefont {J.}~\bibnamefont
  {Kurths}},\ }\href {https://doi.org/10.1017/cbo9780511755743} {\emph
  {\bibinfo {title} {Synchronization: A Universal Concept in Nonlinear
  Sciences}}}\ (\bibinfo  {publisher} {Cambridge University Press},\ \bibinfo
  {year} {2001})\BibitemShut {NoStop}%
\bibitem [{\citenamefont {Jeanneret}\ and\ \citenamefont
  {Benz}(2009)}]{jeanneret2009application}%
  \BibitemOpen
  \bibfield  {author} {\bibinfo {author} {\bibfnamefont {B.}~\bibnamefont
  {Jeanneret}}\ and\ \bibinfo {author} {\bibfnamefont {S.~P.}\ \bibnamefont
  {Benz}},\ }\bibfield  {title} {\bibinfo {title} {Application of the
  {{J}}osephson effect in electrical metrology},\ }\href
  {https://doi.org/10.1140/epjst/e2009-01050-6} {\bibfield  {journal} {\bibinfo
   {journal} {The European Physical Journal Special Topics}\ }\textbf {\bibinfo
  {volume} {172}},\ \bibinfo {pages} {181–206} (\bibinfo {year}
  {2009})}\BibitemShut {NoStop}%
\bibitem [{\citenamefont {Mooij}\ and\ \citenamefont
  {Nazarov}(2006)}]{Mooij2006}%
  \BibitemOpen
  \bibfield  {author} {\bibinfo {author} {\bibfnamefont {J.~E.}\ \bibnamefont
  {Mooij}}\ and\ \bibinfo {author} {\bibfnamefont {Y.~V.}\ \bibnamefont
  {Nazarov}},\ }\bibfield  {title} {\bibinfo {title} {Superconducting nanowires
  as quantum phase-slip junctions},\ }\href {https://doi.org/10.1038/nphys234}
  {\bibfield  {journal} {\bibinfo  {journal} {Nature Physics}\ }\textbf
  {\bibinfo {volume} {2}},\ \bibinfo {pages} {169–172} (\bibinfo {year}
  {2006})}\BibitemShut {NoStop}%
\bibitem [{\citenamefont {Averin}\ \emph {et~al.}(1985)\citenamefont {Averin},
  \citenamefont {Zorin},\ and\ \citenamefont {Likharev}}]{averin1985bloch}%
  \BibitemOpen
  \bibfield  {author} {\bibinfo {author} {\bibfnamefont {D.}~\bibnamefont
  {Averin}}, \bibinfo {author} {\bibfnamefont {A.}~\bibnamefont {Zorin}},\ and\
  \bibinfo {author} {\bibfnamefont {K.}~\bibnamefont {Likharev}},\ }\bibfield
  {title} {\bibinfo {title} {{{B}}loch oscillations in small {{J}}osephson
  junctions},\ }\href
  {http://www.jetp.ras.ru/cgi-bin/e/index/e/61/2/p407?a=list} {\bibfield
  {journal} {\bibinfo  {journal} {Sov. Phys. JETP}\ }\textbf {\bibinfo {volume}
  {61}},\ \bibinfo {pages} {407} (\bibinfo {year} {1985})}\BibitemShut
  {NoStop}%
\bibitem [{\citenamefont {Averin}\ \emph {et~al.}(1990)\citenamefont {Averin},
  \citenamefont {Nazarov},\ and\ \citenamefont {Odintsov}}]{Averin1990}%
  \BibitemOpen
  \bibfield  {author} {\bibinfo {author} {\bibfnamefont {D.}~\bibnamefont
  {Averin}}, \bibinfo {author} {\bibfnamefont {Y.}~\bibnamefont {Nazarov}},\
  and\ \bibinfo {author} {\bibfnamefont {A.}~\bibnamefont {Odintsov}},\
  }\bibfield  {title} {\bibinfo {title} {Incoherent tunneling of the {{C}}ooper
  pairs and magnetic flux quanta in ultrasmall {{J}}osephson junctions},\
  }\href {https://doi.org/10.1016/s0921-4526(09)80058-6} {\bibfield  {journal}
  {\bibinfo  {journal} {Physica B: Condensed Matter}\ }\textbf {\bibinfo
  {volume} {165–166}},\ \bibinfo {pages} {945–946} (\bibinfo {year}
  {1990})}\BibitemShut {NoStop}%
\bibitem [{\citenamefont {Guichard}\ and\ \citenamefont
  {Hekking}(2010)}]{Guichard2010}%
  \BibitemOpen
  \bibfield  {author} {\bibinfo {author} {\bibfnamefont {W.}~\bibnamefont
  {Guichard}}\ and\ \bibinfo {author} {\bibfnamefont {F.~W.~J.}\ \bibnamefont
  {Hekking}},\ }\bibfield  {title} {\bibinfo {title} {Phase-charge duality in
  josephson junction circuits: Role of inertia and effect of microwave
  irradiation},\ }\href {https://doi.org/10.1103/PhysRevB.81.064508} {\bibfield
   {journal} {\bibinfo  {journal} {Phys. Rev. B}\ }\textbf {\bibinfo {volume}
  {81}},\ \bibinfo {pages} {064508} (\bibinfo {year} {2010})}\BibitemShut
  {NoStop}%
\bibitem [{\citenamefont {Kuzmin}\ and\ \citenamefont
  {Haviland}(1991)}]{kuzmin1991observation}%
  \BibitemOpen
  \bibfield  {author} {\bibinfo {author} {\bibfnamefont {L.}~\bibnamefont
  {Kuzmin}}\ and\ \bibinfo {author} {\bibfnamefont {D.}~\bibnamefont
  {Haviland}},\ }\bibfield  {title} {\bibinfo {title} {Observation of the
  {{B}}loch oscillations in an ultrasmall {{J}}osephson junction},\ }\href
  {https://doi.org/10.1103/PhysRevLett.67.2890} {\bibfield  {journal} {\bibinfo
   {journal} {Physical Review Letters}\ }\textbf {\bibinfo {volume} {67}},\
  \bibinfo {pages} {2890} (\bibinfo {year} {1991})}\BibitemShut {NoStop}%
\bibitem [{\citenamefont {Kuzmin}\ \emph {et~al.}(1994)\citenamefont {Kuzmin},
  \citenamefont {Pashkin}, \citenamefont {Zorin},\ and\ \citenamefont
  {Claeson}}]{kuzmin1994linewidth}%
  \BibitemOpen
  \bibfield  {author} {\bibinfo {author} {\bibfnamefont {L.}~\bibnamefont
  {Kuzmin}}, \bibinfo {author} {\bibfnamefont {Y.}~\bibnamefont {Pashkin}},
  \bibinfo {author} {\bibfnamefont {A.}~\bibnamefont {Zorin}},\ and\ \bibinfo
  {author} {\bibfnamefont {T.}~\bibnamefont {Claeson}},\ }\bibfield  {title}
  {\bibinfo {title} {Linewidth of {{B}}loch oscillations in small {{J}}osephson
  junctions},\ }\href {https://doi.org/10.1016/0921-4526(94)90083-3} {\bibfield
   {journal} {\bibinfo  {journal} {Physica B: Condensed Matter}\ }\textbf
  {\bibinfo {volume} {203}},\ \bibinfo {pages} {376–380} (\bibinfo {year}
  {1994})}\BibitemShut {NoStop}%
\bibitem [{\citenamefont {Shaikhaidarov}\ \emph {et~al.}(2022)\citenamefont
  {Shaikhaidarov}, \citenamefont {Kim}, \citenamefont {Dunstan}, \citenamefont
  {Antonov}, \citenamefont {Linzen}, \citenamefont {Ziegler}, \citenamefont
  {Golubev}, \citenamefont {Antonov}, \citenamefont {Il’ichev},\ and\
  \citenamefont {Astafiev}}]{Shaikhaidarov2022}%
  \BibitemOpen
  \bibfield  {author} {\bibinfo {author} {\bibfnamefont {R.~S.}\ \bibnamefont
  {Shaikhaidarov}}, \bibinfo {author} {\bibfnamefont {K.~H.}\ \bibnamefont
  {Kim}}, \bibinfo {author} {\bibfnamefont {J.~W.}\ \bibnamefont {Dunstan}},
  \bibinfo {author} {\bibfnamefont {I.~V.}\ \bibnamefont {Antonov}}, \bibinfo
  {author} {\bibfnamefont {S.}~\bibnamefont {Linzen}}, \bibinfo {author}
  {\bibfnamefont {M.}~\bibnamefont {Ziegler}}, \bibinfo {author} {\bibfnamefont
  {D.~S.}\ \bibnamefont {Golubev}}, \bibinfo {author} {\bibfnamefont {V.~N.}\
  \bibnamefont {Antonov}}, \bibinfo {author} {\bibfnamefont {E.~V.}\
  \bibnamefont {Il’ichev}},\ and\ \bibinfo {author} {\bibfnamefont {O.~V.}\
  \bibnamefont {Astafiev}},\ }\bibfield  {title} {\bibinfo {title} {Quantized
  current steps due to the a.c. coherent quantum phase-slip effect},\ }\href
  {https://doi.org/10.1038/s41586-022-04947-z} {\bibfield  {journal} {\bibinfo
  {journal} {Nature}\ }\textbf {\bibinfo {volume} {608}},\ \bibinfo {pages}
  {45–49} (\bibinfo {year} {2022})}\BibitemShut {NoStop}%
\bibitem [{\citenamefont {Kaap}\ \emph
  {et~al.}(2024{\natexlab{a}})\citenamefont {Kaap}, \citenamefont {Kissling},
  \citenamefont {Gaydamachenko}, \citenamefont {Grünhaupt},\ and\
  \citenamefont {Lotkhov}}]{kaap2024demonstration}%
  \BibitemOpen
  \bibfield  {author} {\bibinfo {author} {\bibfnamefont {F.}~\bibnamefont
  {Kaap}}, \bibinfo {author} {\bibfnamefont {C.}~\bibnamefont {Kissling}},
  \bibinfo {author} {\bibfnamefont {V.}~\bibnamefont {Gaydamachenko}}, \bibinfo
  {author} {\bibfnamefont {L.}~\bibnamefont {Grünhaupt}},\ and\ \bibinfo
  {author} {\bibfnamefont {S.}~\bibnamefont {Lotkhov}},\ }\bibfield  {title}
  {\bibinfo {title} {Demonstration of dual {{S}}hapiro steps in small
  {{J}}osephson junctions},\ }\href
  {https://doi.org/10.1038/s41467-024-53011-z} {\bibfield  {journal} {\bibinfo
  {journal} {Nat. Commun.}\ }\textbf {\bibinfo {volume} {15}},\ \bibinfo
  {pages} {8726} (\bibinfo {year} {2024}{\natexlab{a}})}\BibitemShut {NoStop}%
\bibitem [{\citenamefont {Kaap}\ \emph
  {et~al.}(2024{\natexlab{b}})\citenamefont {Kaap}, \citenamefont {Scheer},
  \citenamefont {Hassler},\ and\ \citenamefont {Lotkhov}}]{Kaap2024}%
  \BibitemOpen
  \bibfield  {author} {\bibinfo {author} {\bibfnamefont {F.}~\bibnamefont
  {Kaap}}, \bibinfo {author} {\bibfnamefont {D.}~\bibnamefont {Scheer}},
  \bibinfo {author} {\bibfnamefont {F.}~\bibnamefont {Hassler}},\ and\ \bibinfo
  {author} {\bibfnamefont {S.}~\bibnamefont {Lotkhov}},\ }\bibfield  {title}
  {\bibinfo {title} {Synchronization of {{B}}loch oscillations in a strongly
  coupled pair of small {{J}}osephson junctions: Evidence for a
  {{S}}hapiro-like current step},\ }\href
  {https://doi.org/10.1103/PhysRevLett.132.027001} {\bibfield  {journal}
  {\bibinfo  {journal} {Phys. Rev. Lett.}\ }\textbf {\bibinfo {volume} {132}},\
  \bibinfo {pages} {027001} (\bibinfo {year} {2024}{\natexlab{b}})}\BibitemShut
  {NoStop}%
\bibitem [{\citenamefont {Crescini}\ \emph {et~al.}(2023)\citenamefont
  {Crescini}, \citenamefont {Cailleaux}, \citenamefont {Guichard},
  \citenamefont {Naud}, \citenamefont {Buisson}, \citenamefont {W.~Murch},\
  and\ \citenamefont {Roch}}]{Crescini2023}%
  \BibitemOpen
  \bibfield  {author} {\bibinfo {author} {\bibfnamefont {N.}~\bibnamefont
  {Crescini}}, \bibinfo {author} {\bibfnamefont {S.}~\bibnamefont {Cailleaux}},
  \bibinfo {author} {\bibfnamefont {W.}~\bibnamefont {Guichard}}, \bibinfo
  {author} {\bibfnamefont {C.}~\bibnamefont {Naud}}, \bibinfo {author}
  {\bibfnamefont {O.}~\bibnamefont {Buisson}}, \bibinfo {author} {\bibfnamefont
  {K.}~\bibnamefont {W.~Murch}},\ and\ \bibinfo {author} {\bibfnamefont
  {N.}~\bibnamefont {Roch}},\ }\bibfield  {title} {\bibinfo {title} {Evidence
  of dual {{S}}hapiro steps in a {{J}}osephson junction array},\ }\href
  {https://doi.org/10.1038/s41567-023-01961-4} {\bibfield  {journal} {\bibinfo
  {journal} {Nature Physics}\ }\textbf {\bibinfo {volume} {19}},\ \bibinfo
  {pages} {851–856} (\bibinfo {year} {2023})}\BibitemShut {NoStop}%
\bibitem [{\citenamefont {Wei\ss{}l}\ \emph {et~al.}(2015)\citenamefont
  {Wei\ss{}l}, \citenamefont {Rastelli}, \citenamefont {Matei}, \citenamefont
  {Pop}, \citenamefont {Buisson}, \citenamefont {Hekking},\ and\ \citenamefont
  {Guichard}}]{Weil2015}%
  \BibitemOpen
  \bibfield  {author} {\bibinfo {author} {\bibfnamefont {T.}~\bibnamefont
  {Wei\ss{}l}}, \bibinfo {author} {\bibfnamefont {G.}~\bibnamefont {Rastelli}},
  \bibinfo {author} {\bibfnamefont {I.}~\bibnamefont {Matei}}, \bibinfo
  {author} {\bibfnamefont {I.~M.}\ \bibnamefont {Pop}}, \bibinfo {author}
  {\bibfnamefont {O.}~\bibnamefont {Buisson}}, \bibinfo {author} {\bibfnamefont
  {F.~W.~J.}\ \bibnamefont {Hekking}},\ and\ \bibinfo {author} {\bibfnamefont
  {W.}~\bibnamefont {Guichard}},\ }\bibfield  {title} {\bibinfo {title}
  {{{B}}loch band dynamics of a {{J}}osephson junction in an inductive
  environment},\ }\href {https://doi.org/10.1103/PhysRevB.91.014507} {\bibfield
   {journal} {\bibinfo  {journal} {Phys. Rev. B}\ }\textbf {\bibinfo {volume}
  {91}},\ \bibinfo {pages} {014507} (\bibinfo {year} {2015})}\BibitemShut
  {NoStop}%
\bibitem [{\citenamefont {Rastelli}\ and\ \citenamefont
  {Pop}(2018)}]{rastelli2018tunable}%
  \BibitemOpen
  \bibfield  {author} {\bibinfo {author} {\bibfnamefont {G.}~\bibnamefont
  {Rastelli}}\ and\ \bibinfo {author} {\bibfnamefont {I.~M.}\ \bibnamefont
  {Pop}},\ }\bibfield  {title} {\bibinfo {title} {Tunable ohmic environment
  using josephson junction chains},\ }\href
  {https://doi.org/10.1103/PhysRevB.97.205429} {\bibfield  {journal} {\bibinfo
  {journal} {Phys. Rev. B}\ }\textbf {\bibinfo {volume} {97}},\ \bibinfo
  {pages} {205429} (\bibinfo {year} {2018})}\BibitemShut {NoStop}%
\bibitem [{\citenamefont {Lolli}\ \emph {et~al.}(2015)\citenamefont {Lolli},
  \citenamefont {Baksic}, \citenamefont {Nagy}, \citenamefont {Manucharyan},\
  and\ \citenamefont {Ciuti}}]{Lolli2015}%
  \BibitemOpen
  \bibfield  {author} {\bibinfo {author} {\bibfnamefont {J.}~\bibnamefont
  {Lolli}}, \bibinfo {author} {\bibfnamefont {A.}~\bibnamefont {Baksic}},
  \bibinfo {author} {\bibfnamefont {D.}~\bibnamefont {Nagy}}, \bibinfo {author}
  {\bibfnamefont {V.~E.}\ \bibnamefont {Manucharyan}},\ and\ \bibinfo {author}
  {\bibfnamefont {C.}~\bibnamefont {Ciuti}},\ }\bibfield  {title} {\bibinfo
  {title} {Ancillary qubit spectroscopy of vacua in cavity and circuit quantum
  electrodynamics},\ }\href {https://doi.org/10.1103/PhysRevLett.114.183601}
  {\bibfield  {journal} {\bibinfo  {journal} {Phys. Rev. Lett.}\ }\textbf
  {\bibinfo {volume} {114}},\ \bibinfo {pages} {183601} (\bibinfo {year}
  {2015})}\BibitemShut {NoStop}%
\bibitem [{\citenamefont {Puertas~Mart{\'\i}nez}\ \emph
  {et~al.}(2019)\citenamefont {Puertas~Mart{\'\i}nez}, \citenamefont
  {L{\'e}ger}, \citenamefont {Gheeraert}, \citenamefont {Dassonneville},
  \citenamefont {Planat}, \citenamefont {Foroughi}, \citenamefont {Krupko},
  \citenamefont {Buisson}, \citenamefont {Naud}, \citenamefont {Hasch-Guichard}
  \emph {et~al.}}]{PuertasMartinez2019Tunable}%
  \BibitemOpen
  \bibfield  {author} {\bibinfo {author} {\bibfnamefont {J.}~\bibnamefont
  {Puertas~Mart{\'\i}nez}}, \bibinfo {author} {\bibfnamefont {S.}~\bibnamefont
  {L{\'e}ger}}, \bibinfo {author} {\bibfnamefont {N.}~\bibnamefont
  {Gheeraert}}, \bibinfo {author} {\bibfnamefont {R.}~\bibnamefont
  {Dassonneville}}, \bibinfo {author} {\bibfnamefont {L.}~\bibnamefont
  {Planat}}, \bibinfo {author} {\bibfnamefont {F.}~\bibnamefont {Foroughi}},
  \bibinfo {author} {\bibfnamefont {Y.}~\bibnamefont {Krupko}}, \bibinfo
  {author} {\bibfnamefont {O.}~\bibnamefont {Buisson}}, \bibinfo {author}
  {\bibfnamefont {C.}~\bibnamefont {Naud}}, \bibinfo {author} {\bibfnamefont
  {W.}~\bibnamefont {Hasch-Guichard}}, \emph {et~al.},\ }\bibfield  {title}
  {\bibinfo {title} {A tunable {{J}}osephson platform to explore many-body
  quantum optics in circuit-{{QED}}},\ }\href
  {https://doi.org/10.1038/s41534-018-0104-0} {\bibfield  {journal} {\bibinfo
  {journal} {npj Quantum Information}\ }\textbf {\bibinfo {volume} {5}},\
  \bibinfo {pages} {19} (\bibinfo {year} {2019})}\BibitemShut {NoStop}%
\bibitem [{\citenamefont {Kuzmin}\ \emph {et~al.}(2019)\citenamefont {Kuzmin},
  \citenamefont {Mehta}, \citenamefont {Grabon}, \citenamefont {Mencia},\ and\
  \citenamefont {Manucharyan}}]{Manucharyan2019npj}%
  \BibitemOpen
  \bibfield  {author} {\bibinfo {author} {\bibfnamefont {R.}~\bibnamefont
  {Kuzmin}}, \bibinfo {author} {\bibfnamefont {N.}~\bibnamefont {Mehta}},
  \bibinfo {author} {\bibfnamefont {N.}~\bibnamefont {Grabon}}, \bibinfo
  {author} {\bibfnamefont {R.}~\bibnamefont {Mencia}},\ and\ \bibinfo {author}
  {\bibfnamefont {V.~E.}\ \bibnamefont {Manucharyan}},\ }\bibfield  {title}
  {\bibinfo {title} {Superstrong coupling in circuit quantum electrodynamics},\
  }\href {https://doi.org/10.1038/s41534-019-0134-2} {\bibfield  {journal}
  {\bibinfo  {journal} {npj Quantum Information}\ }\textbf {\bibinfo {volume}
  {5}},\ \bibinfo {pages} {20} (\bibinfo {year} {2019})}\BibitemShut {NoStop}%
\bibitem [{\citenamefont {Planat}\ \emph {et~al.}(2019)\citenamefont {Planat},
  \citenamefont {Al-Tavil}, \citenamefont {Mart\'{\i}nez}, \citenamefont
  {Dassonneville}, \citenamefont {Foroughi}, \citenamefont {L\'eger},
  \citenamefont {Bharadwaj}, \citenamefont {Delaforce}, \citenamefont
  {Milchakov}, \citenamefont {Naud}, \citenamefont {Buisson}, \citenamefont
  {Hasch-Guichard},\ and\ \citenamefont {Roch}}]{Planat2019}%
  \BibitemOpen
  \bibfield  {author} {\bibinfo {author} {\bibfnamefont {L.}~\bibnamefont
  {Planat}}, \bibinfo {author} {\bibfnamefont {E.}~\bibnamefont {Al-Tavil}},
  \bibinfo {author} {\bibfnamefont {J.~P.}\ \bibnamefont {Mart\'{\i}nez}},
  \bibinfo {author} {\bibfnamefont {R.}~\bibnamefont {Dassonneville}}, \bibinfo
  {author} {\bibfnamefont {F.}~\bibnamefont {Foroughi}}, \bibinfo {author}
  {\bibfnamefont {S.}~\bibnamefont {L\'eger}}, \bibinfo {author} {\bibfnamefont
  {K.}~\bibnamefont {Bharadwaj}}, \bibinfo {author} {\bibfnamefont
  {J.}~\bibnamefont {Delaforce}}, \bibinfo {author} {\bibfnamefont
  {V.}~\bibnamefont {Milchakov}}, \bibinfo {author} {\bibfnamefont
  {C.}~\bibnamefont {Naud}}, \bibinfo {author} {\bibfnamefont {O.}~\bibnamefont
  {Buisson}}, \bibinfo {author} {\bibfnamefont {W.}~\bibnamefont
  {Hasch-Guichard}},\ and\ \bibinfo {author} {\bibfnamefont {N.}~\bibnamefont
  {Roch}},\ }\bibfield  {title} {\bibinfo {title} {Fabrication and
  characterization of aluminum {{SQUID}} transmission lines},\ }\href
  {https://doi.org/10.1103/PhysRevApplied.12.064017} {\bibfield  {journal}
  {\bibinfo  {journal} {Phys. Rev. Appl.}\ }\textbf {\bibinfo {volume} {12}},\
  \bibinfo {pages} {064017} (\bibinfo {year} {2019})}\BibitemShut {NoStop}%
\bibitem [{\citenamefont {Forn-Díaz}\ \emph {et~al.}(2016)\citenamefont
  {Forn-Díaz}, \citenamefont {García-Ripoll}, \citenamefont {Peropadre},
  \citenamefont {Orgiazzi}, \citenamefont {Yurtalan}, \citenamefont
  {Belyansky}, \citenamefont {Wilson},\ and\ \citenamefont
  {Lupascu}}]{FornDaz2016}%
  \BibitemOpen
  \bibfield  {author} {\bibinfo {author} {\bibfnamefont {P.}~\bibnamefont
  {Forn-Díaz}}, \bibinfo {author} {\bibfnamefont {J.~J.}\ \bibnamefont
  {García-Ripoll}}, \bibinfo {author} {\bibfnamefont {B.}~\bibnamefont
  {Peropadre}}, \bibinfo {author} {\bibfnamefont {J.-L.}\ \bibnamefont
  {Orgiazzi}}, \bibinfo {author} {\bibfnamefont {M.~A.}\ \bibnamefont
  {Yurtalan}}, \bibinfo {author} {\bibfnamefont {R.}~\bibnamefont {Belyansky}},
  \bibinfo {author} {\bibfnamefont {C.~M.}\ \bibnamefont {Wilson}},\ and\
  \bibinfo {author} {\bibfnamefont {A.}~\bibnamefont {Lupascu}},\ }\bibfield
  {title} {\bibinfo {title} {Ultrastrong coupling of a single artificial atom
  to an electromagnetic continuum in the nonperturbative regime},\ }\href
  {https://doi.org/10.1038/nphys3905} {\bibfield  {journal} {\bibinfo
  {journal} {Nature Physics}\ }\textbf {\bibinfo {volume} {13}},\ \bibinfo
  {pages} {39–43} (\bibinfo {year} {2016})}\BibitemShut {NoStop}%
\bibitem [{\citenamefont {Kuzmin}\ \emph {et~al.}(2021)\citenamefont {Kuzmin},
  \citenamefont {Grabon}, \citenamefont {Mehta}, \citenamefont {Burshtein},
  \citenamefont {Goldstein}, \citenamefont {Houzet}, \citenamefont {Glazman},\
  and\ \citenamefont {Manucharyan}}]{Kuzmin2021}%
  \BibitemOpen
  \bibfield  {author} {\bibinfo {author} {\bibfnamefont {R.}~\bibnamefont
  {Kuzmin}}, \bibinfo {author} {\bibfnamefont {N.}~\bibnamefont {Grabon}},
  \bibinfo {author} {\bibfnamefont {N.}~\bibnamefont {Mehta}}, \bibinfo
  {author} {\bibfnamefont {A.}~\bibnamefont {Burshtein}}, \bibinfo {author}
  {\bibfnamefont {M.}~\bibnamefont {Goldstein}}, \bibinfo {author}
  {\bibfnamefont {M.}~\bibnamefont {Houzet}}, \bibinfo {author} {\bibfnamefont
  {L.~I.}\ \bibnamefont {Glazman}},\ and\ \bibinfo {author} {\bibfnamefont
  {V.~E.}\ \bibnamefont {Manucharyan}},\ }\bibfield  {title} {\bibinfo {title}
  {Inelastic scattering of a photon by a quantum phase slip},\ }\href
  {https://doi.org/10.1103/PhysRevLett.126.197701} {\bibfield  {journal}
  {\bibinfo  {journal} {Phys. Rev. Lett.}\ }\textbf {\bibinfo {volume} {126}},\
  \bibinfo {pages} {197701} (\bibinfo {year} {2021})}\BibitemShut {NoStop}%
\bibitem [{\citenamefont {Magazzù}\ \emph {et~al.}(2018)\citenamefont
  {Magazzù}, \citenamefont {Forn-Díaz}, \citenamefont {Belyansky},
  \citenamefont {Orgiazzi}, \citenamefont {Yurtalan}, \citenamefont {Otto},
  \citenamefont {Lupascu}, \citenamefont {Wilson},\ and\ \citenamefont
  {Grifoni}}]{Magazz2018}%
  \BibitemOpen
  \bibfield  {author} {\bibinfo {author} {\bibfnamefont {L.}~\bibnamefont
  {Magazzù}}, \bibinfo {author} {\bibfnamefont {P.}~\bibnamefont
  {Forn-Díaz}}, \bibinfo {author} {\bibfnamefont {R.}~\bibnamefont
  {Belyansky}}, \bibinfo {author} {\bibfnamefont {J.-L.}\ \bibnamefont
  {Orgiazzi}}, \bibinfo {author} {\bibfnamefont {M.~A.}\ \bibnamefont
  {Yurtalan}}, \bibinfo {author} {\bibfnamefont {M.~R.}\ \bibnamefont {Otto}},
  \bibinfo {author} {\bibfnamefont {A.}~\bibnamefont {Lupascu}}, \bibinfo
  {author} {\bibfnamefont {C.~M.}\ \bibnamefont {Wilson}},\ and\ \bibinfo
  {author} {\bibfnamefont {M.}~\bibnamefont {Grifoni}},\ }\bibfield  {title}
  {\bibinfo {title} {Probing the strongly driven spin-boson model in a
  superconducting quantum circuit},\ }\href
  {https://doi.org/10.1038/s41467-018-03626-w} {\bibfield  {journal} {\bibinfo
  {journal} {Nature Communications}\ }\textbf {\bibinfo {volume} {9}},\
  \bibinfo {pages} {1403} (\bibinfo {year} {2018})}\BibitemShut {NoStop}%
\bibitem [{\citenamefont {L{\'e}ger}\ \emph {et~al.}(2019)\citenamefont
  {L{\'e}ger}, \citenamefont {Puertas-Mart{\'\i}nez}, \citenamefont
  {Bharadwaj}, \citenamefont {Dassonneville}, \citenamefont {Delaforce},
  \citenamefont {Foroughi}, \citenamefont {Milchakov}, \citenamefont {Planat},
  \citenamefont {Buisson}, \citenamefont {Naud} \emph
  {et~al.}}]{leger2019observation}%
  \BibitemOpen
  \bibfield  {author} {\bibinfo {author} {\bibfnamefont {S.}~\bibnamefont
  {L{\'e}ger}}, \bibinfo {author} {\bibfnamefont {J.}~\bibnamefont
  {Puertas-Mart{\'\i}nez}}, \bibinfo {author} {\bibfnamefont {K.}~\bibnamefont
  {Bharadwaj}}, \bibinfo {author} {\bibfnamefont {R.}~\bibnamefont
  {Dassonneville}}, \bibinfo {author} {\bibfnamefont {J.}~\bibnamefont
  {Delaforce}}, \bibinfo {author} {\bibfnamefont {F.}~\bibnamefont {Foroughi}},
  \bibinfo {author} {\bibfnamefont {V.}~\bibnamefont {Milchakov}}, \bibinfo
  {author} {\bibfnamefont {L.}~\bibnamefont {Planat}}, \bibinfo {author}
  {\bibfnamefont {O.}~\bibnamefont {Buisson}}, \bibinfo {author} {\bibfnamefont
  {C.}~\bibnamefont {Naud}}, \emph {et~al.},\ }\bibfield  {title} {\bibinfo
  {title} {Observation of quantum many-body effects due to zero point
  fluctuations in superconducting circuits},\ }\href
  {https://doi.org/10.1038/s41467-019-13199-x} {\bibfield  {journal} {\bibinfo
  {journal} {Nature Communications}\ }\textbf {\bibinfo {volume} {10}},\
  \bibinfo {pages} {5259} (\bibinfo {year} {2019})}\BibitemShut {NoStop}%
\bibitem [{\citenamefont {Mehta}\ \emph {et~al.}(2022)\citenamefont {Mehta},
  \citenamefont {Ciuti}, \citenamefont {Kuzmin},\ and\ \citenamefont
  {Manucharyan}}]{Manucharyan2022}%
  \BibitemOpen
  \bibfield  {author} {\bibinfo {author} {\bibfnamefont {N.}~\bibnamefont
  {Mehta}}, \bibinfo {author} {\bibfnamefont {C.}~\bibnamefont {Ciuti}},
  \bibinfo {author} {\bibfnamefont {R.}~\bibnamefont {Kuzmin}},\ and\ \bibinfo
  {author} {\bibfnamefont {V.~E.}\ \bibnamefont {Manucharyan}},\ }\bibfield
  {title} {\bibinfo {title} {Theory of strong down-conversion in multi-mode
  cavity and circuit {{QED}}},\ }\href {https://arxiv.org/abs/2210.14681}
  {\bibfield  {journal} {\bibinfo  {journal} {arXiv preprint arXiv:2210.14681}\
  } (\bibinfo {year} {2022})}\BibitemShut {NoStop}%
\bibitem [{\citenamefont {Somoroff}\ \emph {et~al.}(2023)\citenamefont
  {Somoroff}, \citenamefont {Ficheux}, \citenamefont {Mencia}, \citenamefont
  {Xiong}, \citenamefont {Kuzmin},\ and\ \citenamefont
  {Manucharyan}}]{Manucharyan2023}%
  \BibitemOpen
  \bibfield  {author} {\bibinfo {author} {\bibfnamefont {A.}~\bibnamefont
  {Somoroff}}, \bibinfo {author} {\bibfnamefont {Q.}~\bibnamefont {Ficheux}},
  \bibinfo {author} {\bibfnamefont {R.~A.}\ \bibnamefont {Mencia}}, \bibinfo
  {author} {\bibfnamefont {H.}~\bibnamefont {Xiong}}, \bibinfo {author}
  {\bibfnamefont {R.}~\bibnamefont {Kuzmin}},\ and\ \bibinfo {author}
  {\bibfnamefont {V.~E.}\ \bibnamefont {Manucharyan}},\ }\bibfield  {title}
  {\bibinfo {title} {Millisecond coherence in a superconducting qubit},\ }\href
  {https://doi.org/10.1103/PhysRevLett.130.267001} {\bibfield  {journal}
  {\bibinfo  {journal} {Phys. Rev. Lett.}\ }\textbf {\bibinfo {volume} {130}},\
  \bibinfo {pages} {267001} (\bibinfo {year} {2023})}\BibitemShut {NoStop}%
\bibitem [{\citenamefont {Mehta}\ \emph {et~al.}(2023)\citenamefont {Mehta},
  \citenamefont {Kuzmin}, \citenamefont {Ciuti},\ and\ \citenamefont
  {Manucharyan}}]{Mehta2023}%
  \BibitemOpen
  \bibfield  {author} {\bibinfo {author} {\bibfnamefont {N.}~\bibnamefont
  {Mehta}}, \bibinfo {author} {\bibfnamefont {R.}~\bibnamefont {Kuzmin}},
  \bibinfo {author} {\bibfnamefont {C.}~\bibnamefont {Ciuti}},\ and\ \bibinfo
  {author} {\bibfnamefont {V.~E.}\ \bibnamefont {Manucharyan}},\ }\bibfield
  {title} {\bibinfo {title} {Down-conversion of a single photon as a probe of
  many-body localization},\ }\href {https://doi.org/10.1038/s41586-022-05615-y}
  {\bibfield  {journal} {\bibinfo  {journal} {Nature}\ }\textbf {\bibinfo
  {volume} {613}},\ \bibinfo {pages} {650–655} (\bibinfo {year}
  {2023})}\BibitemShut {NoStop}%
\bibitem [{\citenamefont {Kuzmin}\ \emph
  {et~al.}(2023{\natexlab{a}})\citenamefont {Kuzmin}, \citenamefont {Mehta},
  \citenamefont {Grabon},\ and\ \citenamefont {Manucharyan}}]{Kuzmin2023}%
  \BibitemOpen
  \bibfield  {author} {\bibinfo {author} {\bibfnamefont {R.}~\bibnamefont
  {Kuzmin}}, \bibinfo {author} {\bibfnamefont {N.}~\bibnamefont {Mehta}},
  \bibinfo {author} {\bibfnamefont {N.}~\bibnamefont {Grabon}},\ and\ \bibinfo
  {author} {\bibfnamefont {V.~E.}\ \bibnamefont {Manucharyan}},\ }\bibfield
  {title} {\bibinfo {title} {Tuning the inductance of {{J}}osephson junction
  arrays without {{SQUID}}s},\ }\href {https://doi.org/10.1063/5.0171047}
  {\bibfield  {journal} {\bibinfo  {journal} {Applied Physics Letters}\
  }\textbf {\bibinfo {volume} {123}},\ \bibinfo {pages} {182602} (\bibinfo
  {year} {2023}{\natexlab{a}})}\BibitemShut {NoStop}%
\bibitem [{\citenamefont {L{\'e}ger}\ \emph {et~al.}(2023)\citenamefont
  {L{\'e}ger}, \citenamefont {S{\'e}pulcre}, \citenamefont {Fraudet},
  \citenamefont {Buisson}, \citenamefont {Naud}, \citenamefont
  {Hasch-Guichard}, \citenamefont {Florens}, \citenamefont {Snyman},
  \citenamefont {Basko},\ and\ \citenamefont {Roch}}]{leger2023revealing}%
  \BibitemOpen
  \bibfield  {author} {\bibinfo {author} {\bibfnamefont {S.}~\bibnamefont
  {L{\'e}ger}}, \bibinfo {author} {\bibfnamefont {T.}~\bibnamefont
  {S{\'e}pulcre}}, \bibinfo {author} {\bibfnamefont {D.}~\bibnamefont
  {Fraudet}}, \bibinfo {author} {\bibfnamefont {O.}~\bibnamefont {Buisson}},
  \bibinfo {author} {\bibfnamefont {C.}~\bibnamefont {Naud}}, \bibinfo {author}
  {\bibfnamefont {W.}~\bibnamefont {Hasch-Guichard}}, \bibinfo {author}
  {\bibfnamefont {S.}~\bibnamefont {Florens}}, \bibinfo {author} {\bibfnamefont
  {I.}~\bibnamefont {Snyman}}, \bibinfo {author} {\bibfnamefont {D.~M.}\
  \bibnamefont {Basko}},\ and\ \bibinfo {author} {\bibfnamefont
  {N.}~\bibnamefont {Roch}},\ }\bibfield  {title} {\bibinfo {title} {Revealing
  the finite-frequency response of a bosonic quantum impurity},\ }\href
  {https://doi.org/10.21468/scipostphys.14.5.130} {\bibfield  {journal}
  {\bibinfo  {journal} {SciPost Physics}\ }\textbf {\bibinfo {volume} {14}},\
  \bibinfo {pages} {130} (\bibinfo {year} {2023})}\BibitemShut {NoStop}%
\bibitem [{\citenamefont {Kuzmin}\ \emph
  {et~al.}(2023{\natexlab{b}})\citenamefont {Kuzmin}, \citenamefont {Mehta},
  \citenamefont {Grabon}, \citenamefont {Mencia}, \citenamefont {Burshtein},
  \citenamefont {Goldstein},\ and\ \citenamefont
  {Manucharyan}}]{Manucharyan2023arXiv}%
  \BibitemOpen
  \bibfield  {author} {\bibinfo {author} {\bibfnamefont {R.}~\bibnamefont
  {Kuzmin}}, \bibinfo {author} {\bibfnamefont {N.}~\bibnamefont {Mehta}},
  \bibinfo {author} {\bibfnamefont {N.}~\bibnamefont {Grabon}}, \bibinfo
  {author} {\bibfnamefont {R.~A.}\ \bibnamefont {Mencia}}, \bibinfo {author}
  {\bibfnamefont {A.}~\bibnamefont {Burshtein}}, \bibinfo {author}
  {\bibfnamefont {M.}~\bibnamefont {Goldstein}},\ and\ \bibinfo {author}
  {\bibfnamefont {V.~E.}\ \bibnamefont {Manucharyan}},\ }\bibfield  {title}
  {\bibinfo {title} {Observation of the {{S}}chmid-{{B}}ulgadaev dissipative
  quantum phase transition},\ }\href {https://arxiv.org/abs/2304.05806}
  {\bibfield  {journal} {\bibinfo  {journal} {arXiv preprint arXiv:2304.05806}\
  } (\bibinfo {year} {2023}{\natexlab{b}})},\ \Eprint
  {https://arxiv.org/abs/2304.05806} {arXiv:2304.05806 [cond-mat.mes-hall]}
  \BibitemShut {NoStop}%
\bibitem [{\citenamefont {Giacomelli}\ and\ \citenamefont
  {Ciuti}(2024)}]{giacomelli2024emergent}%
  \BibitemOpen
  \bibfield  {author} {\bibinfo {author} {\bibfnamefont {L.}~\bibnamefont
  {Giacomelli}}\ and\ \bibinfo {author} {\bibfnamefont {C.}~\bibnamefont
  {Ciuti}},\ }\bibfield  {title} {\bibinfo {title} {Emergent quantum phase
  transition of a {{J}}osephson junction coupled to a high-impedance multimode
  resonator},\ }\href {https://arxiv.org/abs/2403.04625} {\bibfield  {journal}
  {\bibinfo  {journal} {Nature Communications}\ ,\ \bibinfo {pages} {5455}}
  (\bibinfo {year} {2024})}\BibitemShut {NoStop}%
\bibitem [{\citenamefont {Ashhab}\ \emph {et~al.}(2024)\citenamefont {Ashhab},
  \citenamefont {Ao}, \citenamefont {Yoshihara}, \citenamefont {Lupascu},\ and\
  \citenamefont {Semba}}]{Ashhab2024}%
  \BibitemOpen
  \bibfield  {author} {\bibinfo {author} {\bibfnamefont {S.}~\bibnamefont
  {Ashhab}}, \bibinfo {author} {\bibfnamefont {Z.}~\bibnamefont {Ao}}, \bibinfo
  {author} {\bibfnamefont {F.}~\bibnamefont {Yoshihara}}, \bibinfo {author}
  {\bibfnamefont {A.}~\bibnamefont {Lupascu}},\ and\ \bibinfo {author}
  {\bibfnamefont {K.}~\bibnamefont {Semba}},\ }\bibfield  {title} {\bibinfo
  {title} {High-frequency suppression of inductive coupling between flux qubit
  and transmission line resonator},\ }\href
  {https://doi.org/10.1088/1402-4896/ad2acf} {\bibfield  {journal} {\bibinfo
  {journal} {Physica Scripta}\ }\textbf {\bibinfo {volume} {99}},\ \bibinfo
  {pages} {045116} (\bibinfo {year} {2024})}\BibitemShut {NoStop}%
\bibitem [{\citenamefont {Mukhopadhyay}\ \emph {et~al.}(2023)\citenamefont
  {Mukhopadhyay}, \citenamefont {Senior}, \citenamefont {Saez-Mollejo},
  \citenamefont {Puglia}, \citenamefont {Zemlicka}, \citenamefont {Fink},\ and\
  \citenamefont {Higginbotham}}]{Mukhopadhyay2023}%
  \BibitemOpen
  \bibfield  {author} {\bibinfo {author} {\bibfnamefont {S.}~\bibnamefont
  {Mukhopadhyay}}, \bibinfo {author} {\bibfnamefont {J.}~\bibnamefont
  {Senior}}, \bibinfo {author} {\bibfnamefont {J.}~\bibnamefont
  {Saez-Mollejo}}, \bibinfo {author} {\bibfnamefont {D.}~\bibnamefont
  {Puglia}}, \bibinfo {author} {\bibfnamefont {M.}~\bibnamefont {Zemlicka}},
  \bibinfo {author} {\bibfnamefont {J.~M.}\ \bibnamefont {Fink}},\ and\
  \bibinfo {author} {\bibfnamefont {A.~P.}\ \bibnamefont {Higginbotham}},\
  }\bibfield  {title} {\bibinfo {title} {Superconductivity from a melted
  insulator in {J}osephson junction arrays},\ }\href
  {https://doi.org/10.1038/s41567-023-02161-w} {\bibfield  {journal} {\bibinfo
  {journal} {Nature Physics}\ }\textbf {\bibinfo {volume} {19}},\ \bibinfo
  {pages} {1630–1635} (\bibinfo {year} {2023})}\BibitemShut {NoStop}%
\bibitem [{\citenamefont {Caldeira}\ and\ \citenamefont
  {Leggett}(1981)}]{Caldeira1981}%
  \BibitemOpen
  \bibfield  {author} {\bibinfo {author} {\bibfnamefont {A.~O.}\ \bibnamefont
  {Caldeira}}\ and\ \bibinfo {author} {\bibfnamefont {A.~J.}\ \bibnamefont
  {Leggett}},\ }\bibfield  {title} {\bibinfo {title} {Influence of dissipation
  on quantum tunneling in macroscopic systems},\ }\href
  {https://doi.org/10.1103/physrevlett.46.211} {\bibfield  {journal} {\bibinfo
  {journal} {Physical Review Letters}\ }\textbf {\bibinfo {volume} {46}},\
  \bibinfo {pages} {211–214} (\bibinfo {year} {1981})}\BibitemShut {NoStop}%
\bibitem [{\citenamefont {Weiss}(2012)}]{weiss2012quantum}%
  \BibitemOpen
  \bibfield  {author} {\bibinfo {author} {\bibfnamefont {U.}~\bibnamefont
  {Weiss}},\ }\href {https://doi.org/10.1142/8334} {\emph {\bibinfo {title}
  {Quantum Dissipative Systems}}}\ (\bibinfo  {publisher} {WORLD SCIENTIFIC},\
  \bibinfo {year} {2012})\BibitemShut {NoStop}%
\bibitem [{\citenamefont {Devoret}(1995)}]{devoret1995quantum}%
  \BibitemOpen
  \bibfield  {author} {\bibinfo {author} {\bibfnamefont {M.~H.}\ \bibnamefont
  {Devoret}},\ }\bibfield  {title} {\bibinfo {title} {Quantum fluctuations in
  electrical circuits},\ }\href
  {https://inis.iaea.org/search/search.aspx?orig_q=RN:29063476} {\bibfield
  {journal} {\bibinfo  {journal} {Les Houches, Session LXIII}\ }\textbf
  {\bibinfo {volume} {7}},\ \bibinfo {pages} {133} (\bibinfo {year}
  {1995})}\BibitemShut {NoStop}%
\bibitem [{\citenamefont {Girvin}(2014)}]{girvin2014circuit}%
  \BibitemOpen
  \bibfield  {author} {\bibinfo {author} {\bibfnamefont {S.~M.}\ \bibnamefont
  {Girvin}},\ }\bibinfo {title} {Circuit {{QED}}: superconducting qubits
  coupled to microwave photons},\ in\ \href
  {https://doi.org/10.1093/acprof:oso/9780199681181.003.0003} {\emph {\bibinfo
  {booktitle} {Quantum Machines: Measurement and Control of Engineered Quantum
  Systems}}}\ (\bibinfo  {publisher} {Oxford University PressOxford},\ \bibinfo
  {year} {2014})\ p.\ \bibinfo {pages} {113–256}\BibitemShut {NoStop}%
\bibitem [{\citenamefont {Blais}\ \emph {et~al.}(2021)\citenamefont {Blais},
  \citenamefont {Grimsmo}, \citenamefont {Girvin},\ and\ \citenamefont
  {Wallraff}}]{blais2021circuit}%
  \BibitemOpen
  \bibfield  {author} {\bibinfo {author} {\bibfnamefont {A.}~\bibnamefont
  {Blais}}, \bibinfo {author} {\bibfnamefont {A.~L.}\ \bibnamefont {Grimsmo}},
  \bibinfo {author} {\bibfnamefont {S.~M.}\ \bibnamefont {Girvin}},\ and\
  \bibinfo {author} {\bibfnamefont {A.}~\bibnamefont {Wallraff}},\ }\bibfield
  {title} {\bibinfo {title} {Circuit quantum electrodynamics},\ }\href
  {https://doi.org/10.1103/RevModPhys.93.025005} {\bibfield  {journal}
  {\bibinfo  {journal} {Rev. Mod. Phys.}\ }\textbf {\bibinfo {volume} {93}},\
  \bibinfo {pages} {025005} (\bibinfo {year} {2021})}\BibitemShut {NoStop}%
\bibitem [{\citenamefont {Polkovnikov}(2010)}]{Polkovnikov2010}%
  \BibitemOpen
  \bibfield  {author} {\bibinfo {author} {\bibfnamefont {A.}~\bibnamefont
  {Polkovnikov}},\ }\bibfield  {title} {\bibinfo {title} {Phase space
  representation of quantum dynamics},\ }\href
  {https://doi.org/10.1016/j.aop.2010.02.006} {\bibfield  {journal} {\bibinfo
  {journal} {Annals of Physics}\ }\textbf {\bibinfo {volume} {325}},\ \bibinfo
  {pages} {1790–1852} (\bibinfo {year} {2010})}\BibitemShut {NoStop}%
\bibitem [{\citenamefont {Steel}\ \emph {et~al.}(1998)\citenamefont {Steel},
  \citenamefont {Olsen}, \citenamefont {Plimak}, \citenamefont {Drummond},
  \citenamefont {Tan}, \citenamefont {Collett}, \citenamefont {Walls},\ and\
  \citenamefont {Graham}}]{Steel1998}%
  \BibitemOpen
  \bibfield  {author} {\bibinfo {author} {\bibfnamefont {M.~J.}\ \bibnamefont
  {Steel}}, \bibinfo {author} {\bibfnamefont {M.~K.}\ \bibnamefont {Olsen}},
  \bibinfo {author} {\bibfnamefont {L.~I.}\ \bibnamefont {Plimak}}, \bibinfo
  {author} {\bibfnamefont {P.~D.}\ \bibnamefont {Drummond}}, \bibinfo {author}
  {\bibfnamefont {S.~M.}\ \bibnamefont {Tan}}, \bibinfo {author} {\bibfnamefont
  {M.~J.}\ \bibnamefont {Collett}}, \bibinfo {author} {\bibfnamefont {D.~F.}\
  \bibnamefont {Walls}},\ and\ \bibinfo {author} {\bibfnamefont
  {R.}~\bibnamefont {Graham}},\ }\bibfield  {title} {\bibinfo {title}
  {Dynamical quantum noise in trapped {{B}}ose-{{E}}instein condensates},\
  }\href {https://doi.org/10.1103/physreva.58.4824} {\bibfield  {journal}
  {\bibinfo  {journal} {Physical Review A}\ }\textbf {\bibinfo {volume} {58}},\
  \bibinfo {pages} {4824–4835} (\bibinfo {year} {1998})}\BibitemShut
  {NoStop}%
\bibitem [{\citenamefont {Kautz}\ and\ \citenamefont
  {Martinis}(1990)}]{Martinis1990}%
  \BibitemOpen
  \bibfield  {author} {\bibinfo {author} {\bibfnamefont {R.~L.}\ \bibnamefont
  {Kautz}}\ and\ \bibinfo {author} {\bibfnamefont {J.~M.}\ \bibnamefont
  {Martinis}},\ }\bibfield  {title} {\bibinfo {title} {Noise-affected {{I-V}}
  curves in small hysteretic {{J}}osephson junctions},\ }\href
  {https://doi.org/10.1103/PhysRevB.42.9903} {\bibfield  {journal} {\bibinfo
  {journal} {Phys. Rev. B}\ }\textbf {\bibinfo {volume} {42}},\ \bibinfo
  {pages} {9903} (\bibinfo {year} {1990})}\BibitemShut {NoStop}%
\bibitem [{\citenamefont {Pekola}\ and\ \citenamefont
  {Karimi}(2024)}]{Pekola2024}%
  \BibitemOpen
  \bibfield  {author} {\bibinfo {author} {\bibfnamefont {J.~P.}\ \bibnamefont
  {Pekola}}\ and\ \bibinfo {author} {\bibfnamefont {B.}~\bibnamefont
  {Karimi}},\ }\bibfield  {title} {\bibinfo {title} {Heat bath in a quantum
  circuit},\ }\href {https://doi.org/10.3390/e26050429} {\bibfield  {journal}
  {\bibinfo  {journal} {Entropy}\ }\textbf {\bibinfo {volume} {26}},\ \bibinfo
  {pages} {429} (\bibinfo {year} {2024})},\ \bibinfo {note} {correspondence:
  jukka.pekola@aalto.fi}\BibitemShut {NoStop}%
\bibitem [{\citenamefont {Koch}\ \emph {et~al.}(2007)\citenamefont {Koch},
  \citenamefont {Yu}, \citenamefont {Gambetta}, \citenamefont {Houck},
  \citenamefont {Schuster}, \citenamefont {Majer}, \citenamefont {Blais},
  \citenamefont {Devoret}, \citenamefont {Girvin},\ and\ \citenamefont
  {Schoelkopf}}]{Koch2007}%
  \BibitemOpen
  \bibfield  {author} {\bibinfo {author} {\bibfnamefont {J.}~\bibnamefont
  {Koch}}, \bibinfo {author} {\bibfnamefont {T.~M.}\ \bibnamefont {Yu}},
  \bibinfo {author} {\bibfnamefont {J.}~\bibnamefont {Gambetta}}, \bibinfo
  {author} {\bibfnamefont {A.~A.}\ \bibnamefont {Houck}}, \bibinfo {author}
  {\bibfnamefont {D.~I.}\ \bibnamefont {Schuster}}, \bibinfo {author}
  {\bibfnamefont {J.}~\bibnamefont {Majer}}, \bibinfo {author} {\bibfnamefont
  {A.}~\bibnamefont {Blais}}, \bibinfo {author} {\bibfnamefont {M.~H.}\
  \bibnamefont {Devoret}}, \bibinfo {author} {\bibfnamefont {S.~M.}\
  \bibnamefont {Girvin}},\ and\ \bibinfo {author} {\bibfnamefont {R.~J.}\
  \bibnamefont {Schoelkopf}},\ }\bibfield  {title} {\bibinfo {title}
  {Charge-insensitive qubit design derived from the {C}ooper pair box},\ }\href
  {https://doi.org/10.1103/PhysRevA.76.042319} {\bibfield  {journal} {\bibinfo
  {journal} {Phys. Rev. A}\ }\textbf {\bibinfo {volume} {76}},\ \bibinfo
  {pages} {042319} (\bibinfo {year} {2007})}\BibitemShut {NoStop}%
\bibitem [{\citenamefont {Jones}\ and\ \citenamefont
  {March}(2003)}]{Jones2003-zy}%
  \BibitemOpen
  \bibfield  {author} {\bibinfo {author} {\bibfnamefont {W.}~\bibnamefont
  {Jones}}\ and\ \bibinfo {author} {\bibfnamefont {N.~H.}\ \bibnamefont
  {March}},\ }\href@noop {} {\emph {\bibinfo {title} {Theoretical solid state
  physics: Perfect lattices in equilibrium v. 1}}},\ Dover Books on Physics\
  (\bibinfo  {publisher} {Dover Publications},\ \bibinfo {address} {Mineola,
  NY},\ \bibinfo {year} {2003})\BibitemShut {NoStop}%
\bibitem [{\citenamefont {Blount}(1962)}]{Blount1962}%
  \BibitemOpen
  \bibfield  {author} {\bibinfo {author} {\bibfnamefont {E.}~\bibnamefont
  {Blount}},\ }\bibinfo {title} {Formalisms of band theory},\ in\ \href
  {https://doi.org/10.1016/s0081-1947(08)60459-2} {\emph {\bibinfo {booktitle}
  {Solid State Physics}}}\ (\bibinfo  {publisher} {Elsevier},\ \bibinfo {year}
  {1962})\ p.\ \bibinfo {pages} {305–373}\BibitemShut {NoStop}%
\bibitem [{\citenamefont {Koch}\ \emph {et~al.}(2009)\citenamefont {Koch},
  \citenamefont {Manucharyan}, \citenamefont {Devoret},\ and\ \citenamefont
  {Glazman}}]{Koch2009}%
  \BibitemOpen
  \bibfield  {author} {\bibinfo {author} {\bibfnamefont {J.}~\bibnamefont
  {Koch}}, \bibinfo {author} {\bibfnamefont {V.}~\bibnamefont {Manucharyan}},
  \bibinfo {author} {\bibfnamefont {M.~H.}\ \bibnamefont {Devoret}},\ and\
  \bibinfo {author} {\bibfnamefont {L.~I.}\ \bibnamefont {Glazman}},\
  }\bibfield  {title} {\bibinfo {title} {Charging effects in the inductively
  shunted {{{{J}}osephson}} junction},\ }\href
  {https://doi.org/10.1103/PhysRevLett.103.217004} {\bibfield  {journal}
  {\bibinfo  {journal} {Phys. Rev. Lett.}\ }\textbf {\bibinfo {volume} {103}},\
  \bibinfo {pages} {217004} (\bibinfo {year} {2009})}\BibitemShut {NoStop}%
\bibitem [{\citenamefont {Pechenezhskiy}\ \emph {et~al.}(2020)\citenamefont
  {Pechenezhskiy}, \citenamefont {Mencia}, \citenamefont {Nguyen},
  \citenamefont {Lin},\ and\ \citenamefont {Manucharyan}}]{Pechenezhskiy2020}%
  \BibitemOpen
  \bibfield  {author} {\bibinfo {author} {\bibfnamefont {I.~V.}\ \bibnamefont
  {Pechenezhskiy}}, \bibinfo {author} {\bibfnamefont {R.~A.}\ \bibnamefont
  {Mencia}}, \bibinfo {author} {\bibfnamefont {L.~B.}\ \bibnamefont {Nguyen}},
  \bibinfo {author} {\bibfnamefont {Y.-H.}\ \bibnamefont {Lin}},\ and\ \bibinfo
  {author} {\bibfnamefont {V.~E.}\ \bibnamefont {Manucharyan}},\ }\bibfield
  {title} {\bibinfo {title} {The superconducting quasicharge qubit},\ }\href
  {https://doi.org/10.1038/s41586-020-2687-9} {\bibfield  {journal} {\bibinfo
  {journal} {Nature}\ }\textbf {\bibinfo {volume} {585}},\ \bibinfo {pages}
  {368–371} (\bibinfo {year} {2020})}\BibitemShut {NoStop}%
\bibitem [{\citenamefont {Gardiner}\ and\ \citenamefont
  {Zoller}(2004)}]{gardiner2004quantum}%
  \BibitemOpen
  \bibfield  {author} {\bibinfo {author} {\bibfnamefont {C.}~\bibnamefont
  {Gardiner}}\ and\ \bibinfo {author} {\bibfnamefont {P.}~\bibnamefont
  {Zoller}},\ }\href {https://www.springer.com/gp/book/9783540223016} {\emph
  {\bibinfo {title} {{Quantum noise: a handbook of Markovian and non-Markovian
  quantum stochastic methods with applications to quantum optics}}}}\ (\bibinfo
   {publisher} {Springer Science \& Business Media},\ \bibinfo {year}
  {2004})\BibitemShut {NoStop}%
\bibitem [{\citenamefont {Sinatra}\ \emph {et~al.}(2002)\citenamefont
  {Sinatra}, \citenamefont {Lobo},\ and\ \citenamefont {Castin}}]{Sinatra2002}%
  \BibitemOpen
  \bibfield  {author} {\bibinfo {author} {\bibfnamefont {A.}~\bibnamefont
  {Sinatra}}, \bibinfo {author} {\bibfnamefont {C.}~\bibnamefont {Lobo}},\ and\
  \bibinfo {author} {\bibfnamefont {Y.}~\bibnamefont {Castin}},\ }\bibfield
  {title} {\bibinfo {title} {The truncated {{W}}igner method for
  {{{{B}}ose}}-condensed gases: Limits of validity and applications},\ }\href
  {https://doi.org/10.1088/0953-4075/35/17/301} {\bibfield  {journal} {\bibinfo
   {journal} {Journal of Physics B: Atomic, Molecular and Optical Physics}\
  }\textbf {\bibinfo {volume} {35}},\ \bibinfo {pages} {3599} (\bibinfo {year}
  {2002})}\BibitemShut {NoStop}%
\bibitem [{\citenamefont {Blakie†}\ \emph {et~al.}(2008)\citenamefont
  {Blakie†}, \citenamefont {Bradley†}, \citenamefont {Davis}, \citenamefont
  {Ballagh},\ and\ \citenamefont {Gardiner}}]{blakie2008dynamics}%
  \BibitemOpen
  \bibfield  {author} {\bibinfo {author} {\bibfnamefont {P.}~\bibnamefont
  {Blakie†}}, \bibinfo {author} {\bibfnamefont {A.}~\bibnamefont
  {Bradley†}}, \bibinfo {author} {\bibfnamefont {M.}~\bibnamefont {Davis}},
  \bibinfo {author} {\bibfnamefont {R.}~\bibnamefont {Ballagh}},\ and\ \bibinfo
  {author} {\bibfnamefont {C.}~\bibnamefont {Gardiner}},\ }\bibfield  {title}
  {\bibinfo {title} {Dynamics and statistical mechanics of ultra-cold {B}ose
  gases using c-field techniques},\ }\href
  {https://doi.org/10.1080/00018730802564254} {\bibfield  {journal} {\bibinfo
  {journal} {Advances in Physics}\ }\textbf {\bibinfo {volume} {57}},\ \bibinfo
  {pages} {363–455} (\bibinfo {year} {2008})}\BibitemShut {NoStop}%
\bibitem [{\citenamefont {Carusotto}\ and\ \citenamefont
  {Ciuti}(2013)}]{carusotto2013quantum}%
  \BibitemOpen
  \bibfield  {author} {\bibinfo {author} {\bibfnamefont {I.}~\bibnamefont
  {Carusotto}}\ and\ \bibinfo {author} {\bibfnamefont {C.}~\bibnamefont
  {Ciuti}},\ }\bibfield  {title} {\bibinfo {title} {Quantum fluids of light},\
  }\href {https://doi.org/10.1103/RevModPhys.85.299} {\bibfield  {journal}
  {\bibinfo  {journal} {Reviews of Modern Physics}\ }\textbf {\bibinfo {volume}
  {85}},\ \bibinfo {pages} {299} (\bibinfo {year} {2013})}\BibitemShut
  {NoStop}%
\bibitem [{\citenamefont {Remez}\ \emph {et~al.}(2024)\citenamefont {Remez},
  \citenamefont {Kurilovich}, \citenamefont {Rieger},\ and\ \citenamefont
  {Glazman}}]{remez2024bloch}%
  \BibitemOpen
  \bibfield  {author} {\bibinfo {author} {\bibfnamefont {B.}~\bibnamefont
  {Remez}}, \bibinfo {author} {\bibfnamefont {V.~D.}\ \bibnamefont
  {Kurilovich}}, \bibinfo {author} {\bibfnamefont {M.}~\bibnamefont {Rieger}},\
  and\ \bibinfo {author} {\bibfnamefont {L.~I.}\ \bibnamefont {Glazman}},\
  }\bibfield  {title} {\bibinfo {title} {{{B}}loch oscillations in a transmon
  embedded in a resonant electromagnetic environment},\ }\href
  {https://doi.org/10.1103/PhysRevB.110.054508} {\bibfield  {journal} {\bibinfo
   {journal} {Phys. Rev. B}\ }\textbf {\bibinfo {volume} {110}},\ \bibinfo
  {pages} {054508} (\bibinfo {year} {2024})}\BibitemShut {NoStop}%
\bibitem [{\citenamefont {Kurilovich}\ \emph {et~al.}(2024)\citenamefont
  {Kurilovich}, \citenamefont {Remez},\ and\ \citenamefont
  {Glazman}}]{kurilovich2024quantum}%
  \BibitemOpen
  \bibfield  {author} {\bibinfo {author} {\bibfnamefont {V.~D.}\ \bibnamefont
  {Kurilovich}}, \bibinfo {author} {\bibfnamefont {B.}~\bibnamefont {Remez}},\
  and\ \bibinfo {author} {\bibfnamefont {L.~I.}\ \bibnamefont {Glazman}},\
  }\bibfield  {title} {\bibinfo {title} {Quantum theory of {{B}}loch
  oscillations in a resistively shunted transmon},\ }\href
  {https://arxiv.org/abs/2403.04624} {\bibfield  {journal} {\bibinfo  {journal}
  {arXiv preprint arXiv:2403.04624}\ } (\bibinfo {year} {2024})},\ \Eprint
  {https://arxiv.org/abs/2403.04624} {"arXiv":2403.04624 [cond-mat.mes-hall]}
  \BibitemShut {NoStop}%
\end{thebibliography}%



\end{document}